\documentclass[11pt]{elsarticle}

\pdfoutput=1

\makeatletter
\def\ps@pprintTitle{%
  \let\@oddhead\@empty
  \let\@evenhead\@empty
  \let\@oddfoot\@empty
  \let\@evenfoot\@oddfoot
}
\makeatother

\usepackage{url}
\usepackage{breakurl}
\usepackage[breaklinks,
            colorlinks = true,
            linkcolor = blue,
            urlcolor  = blue,
            citecolor = blue,
            anchorcolor = blue]{hyperref}

\usepackage{lineno}
\usepackage{textcomp}
\usepackage{graphicx}
\usepackage[margin=1.25in]{geometry}
\usepackage[usenames,dvipsnames]{color}
\usepackage{fancyhdr}
\fancypagestyle{plain}{%
  \fancyhf{}%
  \fancyhead[C]{}
  \fancyfoot[C]{\thepage}
}

\fancypagestyle{empty}{%
  \fancyhf{}%
  \fancyhead[C]{{\it Activity report on the African School of Physics, April 15--19 and July 7--21, 2024}}
  \fancyfoot[C]{\thepage}
}
\begin{document}

\begin{frontmatter}

%\pubblock

\title{Activity Report on the Eighth African School of Fundamental Physics and Applications (ASP2024)}

\author[add1]{K\'et\'evi A. Assamagan\corref{cor1}}
\ead{ketevi@bnl.gov}
\author[add1]{Mounia Laassiri\corref{cor1}}
\ead{mlaassiri@bnl.gov}
\author[add2]{Bobby Acharya}
\author[add3]{Christine Darve}
\author[add4]{Fernando Ferroni}
\author[add5]{Mohamed Chabab}
\author[add6]{Farida Fassi}
\author[add7]{Kenneth Cecire}
\author[add8]{Julia Ann Gray}

\cortext[cor1]{Contacts}

\address[add1]{Brookhaven National Laboratory, USA}
\address[add2]{ICTP, Italy, and King's College London, UK}
\address[add3]{European Spallation Source, Sweden}
\address[add4]{INFN-GSSI, Italy}
\address[add5]{Cadi Ayyad University, Marrakesh, Morocco}
\address[add6]{Mohammed V University, Rabat, Morocco}
\address[add7]{University of Notre Dame, IN, USA}
\address[add8]{ASP International Advisory Committee, Switzerland}

\begin{abstract}
\noindent 
The African School of Fundamental Physics and Applications, also known as the African School of Physics (ASP), was initiated in 2010, as a three-week biennial event, to offer additional training in fundamental and applied physics to African students with a minimum of three-year university education. Since its inception, ASP has grown to be much more than a school. ASP has become a series of activities and events to support academic development of African students, teachers and faculties. We report on the eighth African School of Physics, ASP2024, organized in Morocco, on April 15--19 and July 7--21, 2024. ASP2024 included programs for university students, high school teachers and high school pupils.
\end{abstract}


\end{frontmatter}

%\linenumbers
%

%\def\thefootnote{\fnsymbol{footnote}}
%\setcounter{footnote}{0}

%\newpage

\section{Introduction}
\label{sec:intro}

The African School of Physics is a collection of activities to support academic growths of African students. One activity is a three-week biennial event organized in different African countries---this event consists of a 2-week intensive school, complemented with a one-week African Conference on Fundamental and Applied Physics (ACP)~\cite{ASP2021-reports, ASP, ASP-reports, asp2018, ASP2022}.  The host country of the next biennial event is selected two and half years in advance through a bidding process. In December 2017, Morocco was selected to host ASP2020 at Cadi Ayyad University in Marrakesh.  However, because of the COVID-19 pandemic, ASP2020 was cancelled and held online in 2021. In December 2019, South Africa was selected to host ASP2022~\cite{ASP2022}. Morocco was then reconsidered for ASP2024, and Togo was selected to host ACP2025~\cite{ACP2025}.

In this paper, we present the activity report of ASP2024. In Section~\ref{sec:prog}, we review the scientific program and discuss the supports received in Section~\ref{sec:sup}. We present the profiles of the participants and the expenditures in Sections~\ref{sec:prof} and~\ref{sec:exp} respectively. Feedback from participants are presented in Section~\ref{sec:feed}. Outlook and conclusions are offered in Sections~\ref{sec:out} and~\ref{sec:conc}.

\section{Scientific Program}
\label{sec:prog}

As mentioned in Section~\ref{sec:intro}, The African School of Physics has evolved well beyond three-week biennial engagements. For broader participation in fundamental fields and related applications, the scientific program includes the major physics areas of interest in Africa, as defined by the African Physical Society (AfPS)~\cite{AfPS}:
\begin{itemize}
   \item Particles and related applications: nuclear physics, particle physics, medical physics, (particle)astrophysics \& cosmology, fluid \& plasma physics, complex systems;
   \item Light sources and their applications: light sources, condensed matter \& materials physics, atomic \& molecular physics, optics \& photonics, physics of earth, biophysics;
   \item Cross-cutting fields: accelerator physics, computing, instrumentation \& detectors.
\end{itemize}
Topics in quantum computing \& quantum information and machine learning \& artificial intelligence are also on the agenda. Furthermore, the ASP program includes the fields of societal engagements, namely: topics related to physics education, community engagement, women in physics, early career physicists and engagements with African policymakers in research and education. Representative details on the scientific program are presented in Ref.~\cite{acp2021}, and in the references therein.

The scientific program was arranged in three distinct components for high school pupils (learners), university students, high school teachers. The learners' program took place on April 15-19, 2024. During that period, we organized outreach events for over a thousand learners from several high schools in the region of Marrakesh~\cite{ReportLearners2024}. 

The program for university students was set up for the entire duration of the school, July 7--21, 2024~\cite{Students2024}. The program for high school teachers was in parallel to the students’ program, during the period of July 8--12, 2024~\cite{Teachers2024}. 

On July 13, 2024, we organized a forum where we carried out---among other topics---panel discussions on the "road to critical skill development in Africa" and "physics outreach efforts by ASP alumni"~\cite{Forum2024}. 

%Details of the scientific agenda can be found in Ref.~\cite{ScientificProgram}.
An article released by the ICTP on ASP2024 is available in Ref.~\cite{ICTP-ASP2024}.

\subsection{Scientific program for learners}
\begin{figure}[!htb]
 \begin{center}
  \includegraphics[width=\textwidth]{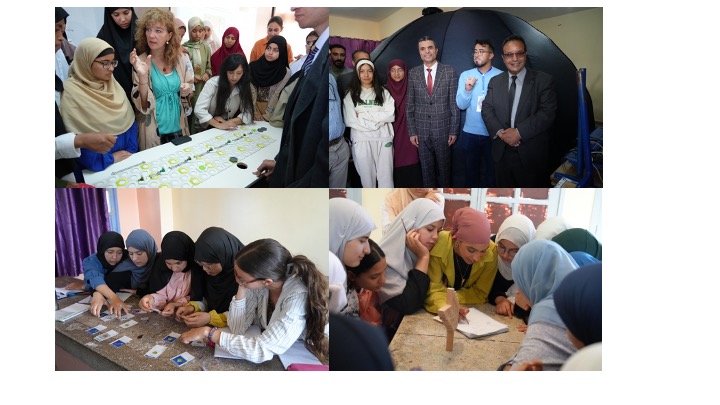}
   \caption{Engagement with learners during ASP2024. The pictures show: Gauss Cannon activities (top left), Regional education authorities, astrophysics instructor and learners in front of a makeshift planetarium (top right), a group of learners working together to order the elementary particles (bottom left), and a discussion on the mathematical formalism of the concept of center-of-mass and the condition of stability of the 15-block Jenga cantilever that the learners succeeded in building (bottom right).}
    \label{fig:learners}
  \end{center}
\end{figure}

The objective for the scientific program for high school pupils is to encourage them to develop and maintain interest in physics. Over one thousand learners from many high schools from the region of Marrakesh, Morocco, attended the events in the period of April 15--19, 2024. Figure~\ref{fig:learners} shows physics outreach sessions with high school learners.
The period of April 15--19, 2024, was chosen to maximize learner's attendance while they are still in school. Five high schools served as the venues, and on different days in April 15--19, 2024, learners from designated high schools converged to a venue. The program was repeated over five days at different venues for different groups of high schools and learners. The participating high schools and learners were selected by the regional education authorities and the school officials. Learners were exposed to physics lectures, demonstrations and hands-on activities in accelerator physics, astrophysics, particle physics, and detectors \& instrumentation. Details on the ASP2024 learners program are documented in Refs.~\cite{ReportLearners2024, Learners2024}.

\subsection{Scientific program for students}
We followed the same organizational structure adopted for ASP2022 and described in Ref.~\cite{ASP2022}. 
\begin{figure}[!htbp]
 \begin{center}
  \includegraphics[width=\textwidth]{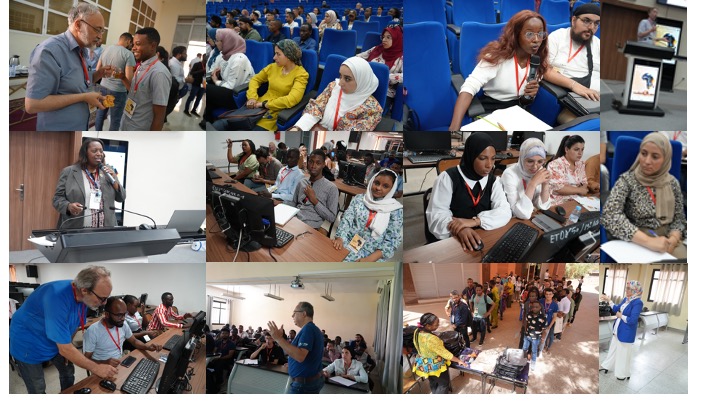}
   \caption{Interactions between students and lecturers during ASP2024.}
    \label{fig:students}
  \end{center}
\end{figure}
Topics of general interests were arranged in plenary sessions during the first week of the school. These were supported by advanced topics developed in parallel sessions. Daily activities started and ended with topics of general interests in plenary sessions. In-between, were combinations of plenary and parallel sessions, to allow for coverage of diverse topics of interest to participants---as presented in Section~\ref{sec:intro}. In the parallel activities, students selected topics that best support the academic majors. Figure~\ref{fig:students} shows snapshots of students' activities.

\subsection{Scientific program for high school teachers}
\begin{figure}[!htbp]
 \begin{center}
  \includegraphics[width=\textwidth]{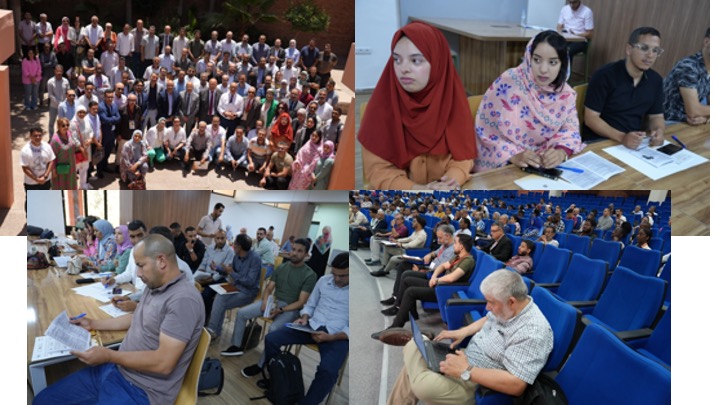}
   \caption{High school teachers during ASP2024.}
    \label{fig:teachers}
  \end{center}
\end{figure}

The objective of the teachers' program is to help in their planning and delivery of physics instructions. The program for teachers took place during the week of July 8--12, 2024; it consisted of physics plenary lectures complemented by three hands-on activities. Teachers received lectures in mechanics, optics and photonics, and thermodynamics. The hands-on sessions included workshops on the Internet of Things, Particle Physics, and Integration of Scientific Computing into Mathematics and Science Classes. Details on the scientific program for high school teachers are in Ref.~\cite{Teachers2024}. A group of 80 Moroccan teachers participated; these were selected nationally by the Ministry of National Education from nearly 200 applications received from 12 regional academies. A joint committee formed by senior officials of the Ministry of National Education and the Local Organizing Committee (LOC) chairs was appointed to review the teachers' program and its objectives, define the selection criteria for teachers, and work on the logistics and financial support needed for a successful program.
Figure~\ref{fig:teachers} shows high school teachers during ASP2024.

\subsection{ASP Forum}
\begin{figure}[!htb]
 \begin{center}
  \includegraphics[width=\textwidth]{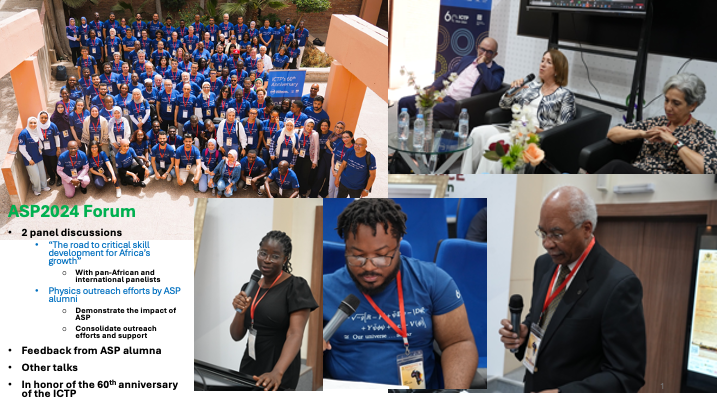}
   \caption{ASP forum for discussion with policymakers on capacity development and retention in Africa. In honor of the sixtieth anniversary celebration of the ICTP, forum participants wore T-shirt donated by the ICTP.}
    \label{fig:forumEvent}
  \end{center}
\end{figure}

With the ASP Forum, we aim to develop or strengthen strategic engagements with African policymakers in physics education and research, for ASP to better serve the interests of the African community. For ASP2024, we organized two panel discussions at the forum. The first one on "the road to critical skills development for Africa growth". Expert panelists from various regions of Africa spoke to the issue in person or online. The second panel discussed focused on physics outreach efforts initiated by or with significant contribution from ASP alumni. Leading to the discussion, four outreach efforts were presented as shown in Ref.~\cite{Forum2024}. Other presentations at the forum included "The vision and plan of the Hassan II Academy of Morocco" and a special on "The Long Crisis of Black Masculinity in Racial Capitalism" for socioeconomic impact on education and career development. The agenda of the ASP2024 Forum can be in Ref.~\cite{Forum2024}. At the conclusion of the forum, we enjoyed a gala dinner. Figure~\ref{fig:forumEvent} shows participants’ feedback on the ASP Forum on July 13, 2024.
\section{Support}
\label{sec:sup}

\begin{figure}[!htbp]
 \begin{center}
  \includegraphics[width=\textwidth]{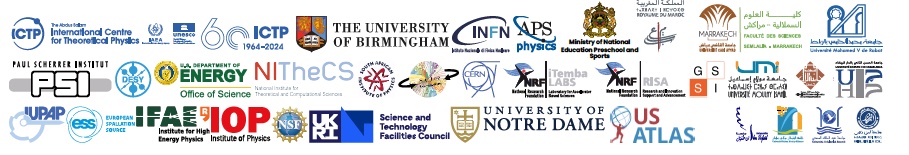}
   \caption{Logos of the institutes that supported ASP2024 financially or in-kind. A fraction of the support was earmarked for ACP2025 as discussed in Section~\ref{sec:exp}. BNL, FNAL, ANL and the MacDonald Institute also contributed to the ASP2024 and ACP2025 budget.}
    \label{fig:logos}
  \end{center}
\end{figure}

We received financial and in-kind support from various institutes whose logos are shown in Figure~\ref{fig:logos}. In addition, there were contributions from Brookhaven National Laboratory (BNL), Fermi National Laboratory (FNAL), Argonne National Laboratory (ANL), and the MacDonald Institute. The support received allowed full coverage of the expenses associated with the event as presented in Section~\ref{sec:exp}. 

\begin{figure}[!htbp]
 \begin{center}
  \includegraphics[width=\textwidth]{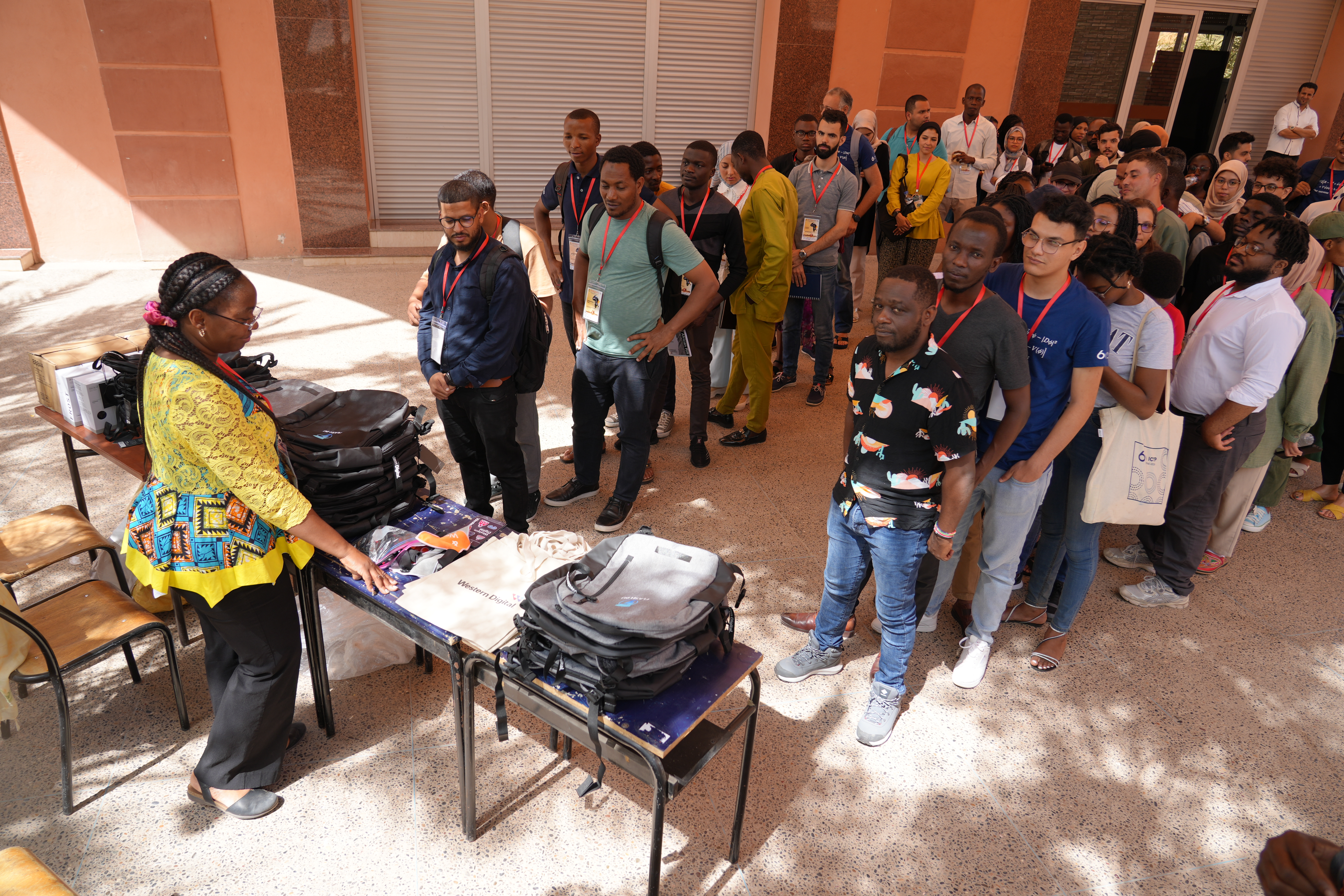}
   \caption{Swag donation, to ASP2024, from Western Digital.}
    \label{fig:swag}
  \end{center}
\end{figure}

Western Digital supported ASP2024 with donation of backpacks, handbags, T-shirts, 16GB-flash drives, tumblers, coolers, card holders, stickers, and screen wipes; the swag was distributed to participants during the event, as shown in Figure~\ref{fig:swag}.

Moroccan students were supported in part by the Higher Council of Education, Training, and Scientific Research (CSEFRS) under CSEFRS -- University Mohammed V Cooperative Agreement. The Ministry of Education covered full room and board for the eighty high school teachers. The hosting faculty provided airport transfers in Marrakesh for the international students. Cadi Ayyad University provided in-kind logistical and technical support. Some lecturers received travel coverage from their institutes.

\section{Participant profiles}
\label{sec:prof}
For the high school outreach program, over one thousand high school learners---of the tenth to the twelfth grades, from many high schools, some of whom are shown in Figure~\ref{fig:learners}---participated, with an average of 200 pupils per day during the period of April 15--19, 2024. The time of the high school outreach was chosen to maximize learners' attendance since they were still in school during that period. Five high schools were selected as the venues were selected learners congregated. The activities were repeated daily, for different learners, at a different venue, over the period of 5 days. the participating learners and high schools were selected by the education authorities for the region of Marrakesh.

Eighty high school teachers were selected by the Moroccan Ministry of Basic Education thru a national call and application process, Figure~\ref{fig:teachers} was taken with the participating teachers on the first day of the event.  

\begin{figure}[!htbp]
 \begin{center}
  \includegraphics[width=\textwidth]{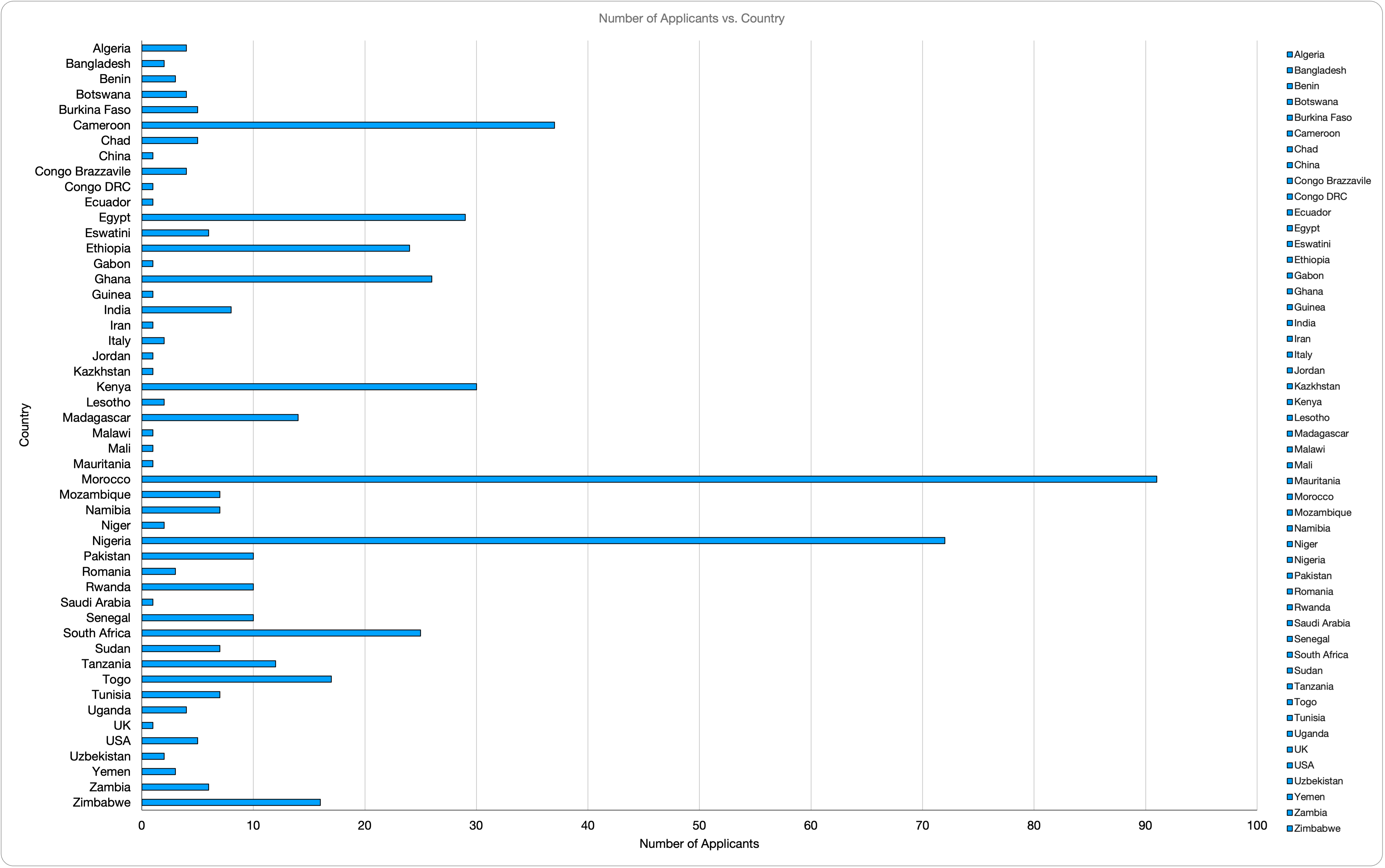}
   \caption{Distribution of ASP2024 student applications as a function of their countries of citizenship.}
    \label{fig:applications}
  \end{center}
\end{figure}

\begin{figure}[!htbp]
 \begin{center}
  \includegraphics[width=\textwidth]{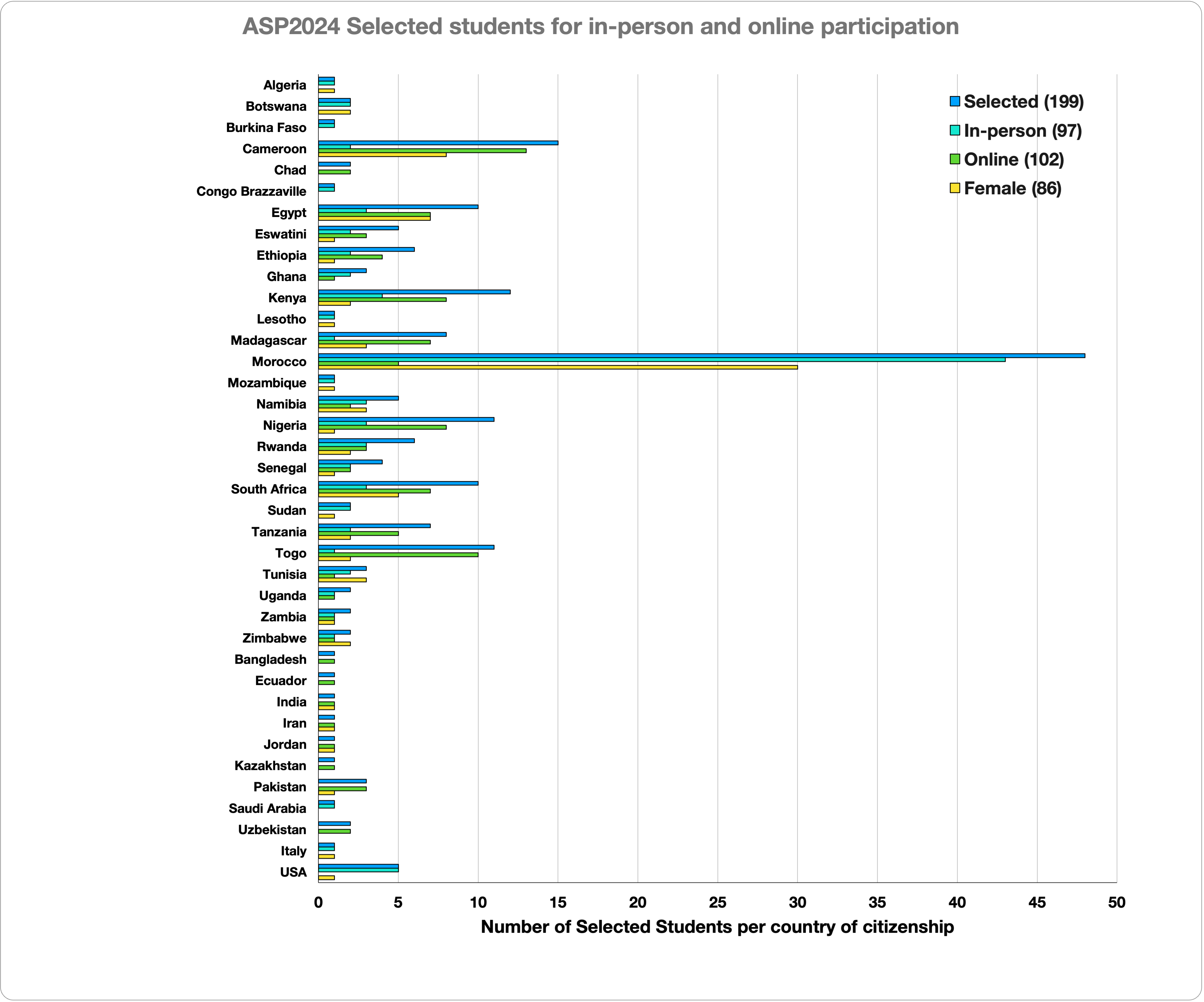}
   \caption{Distribution of selected students for ASP2024 as a function of country of citizenship.}
    \label{fig:selections}
  \end{center}
\end{figure}

\begin{figure}[!htb]
 \begin{center}
  \includegraphics[width=\textwidth]{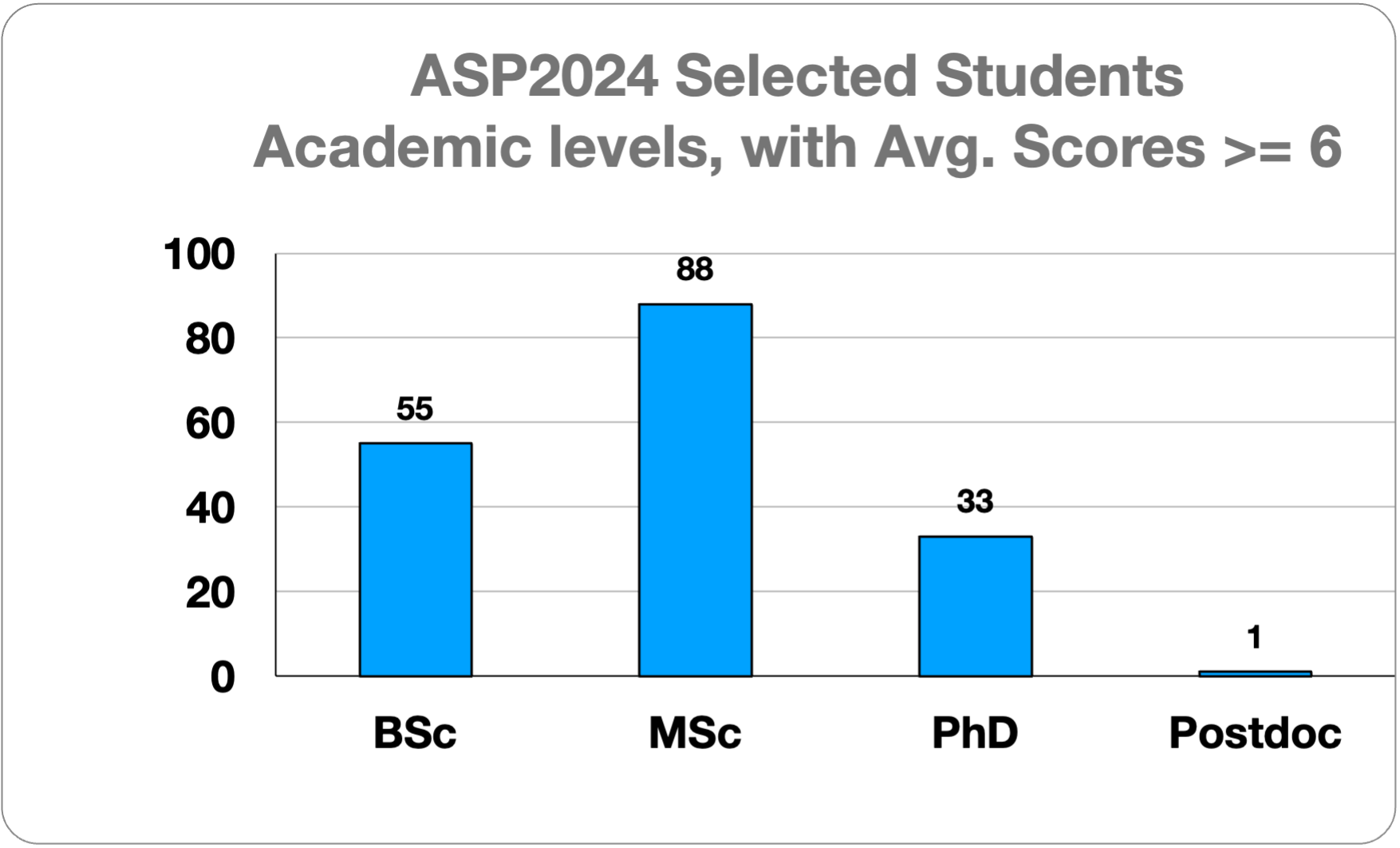}
   \caption{Academic levels of the selected students at ASP2024, at the time of their applications.}
    \label{fig:degrees}
  \end{center}
\end{figure}

\begin{figure}[!htb]
 \begin{center}
  \includegraphics[width=\textwidth]{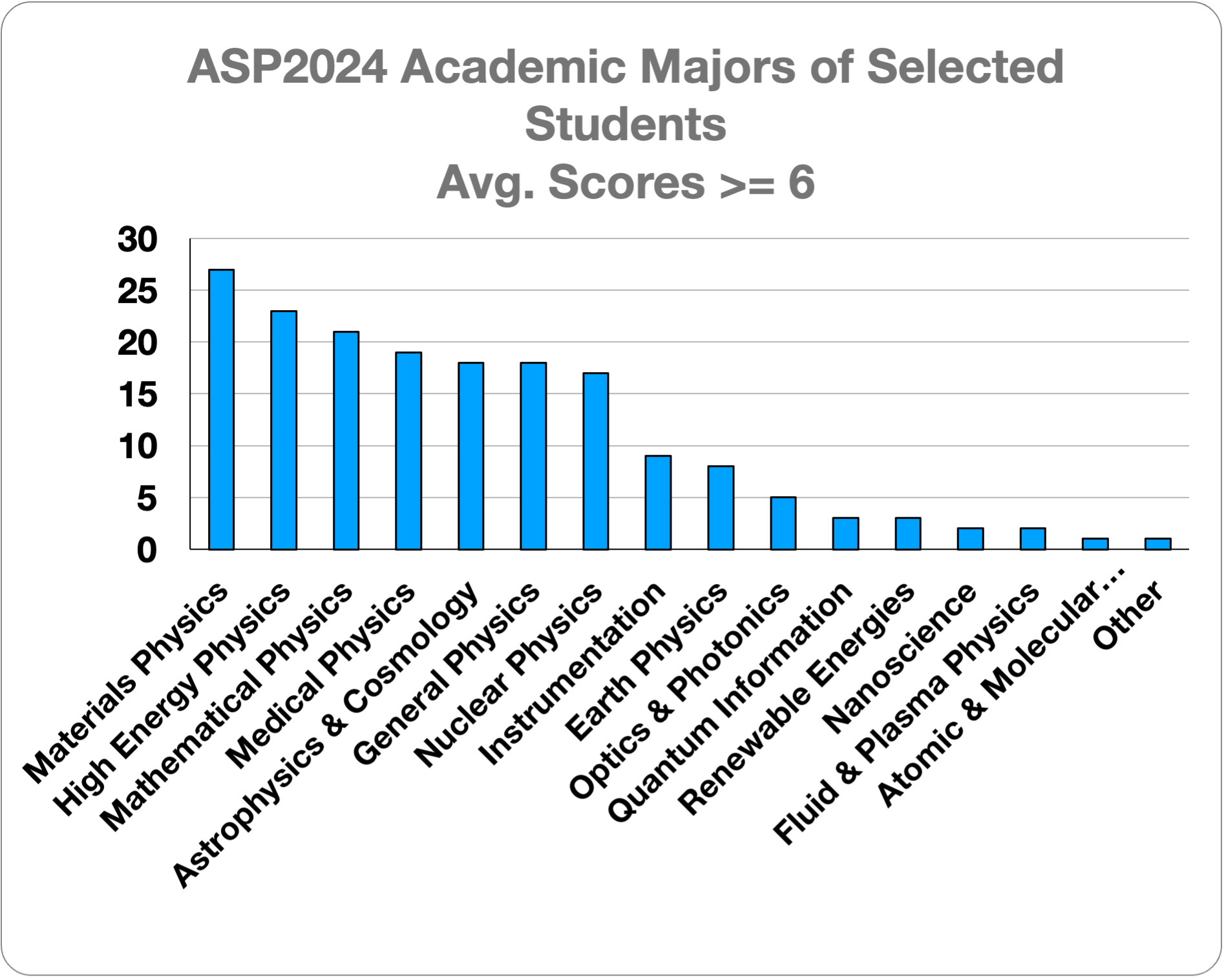}
   \caption{Academic concentrations of the selected students at ASP2024, at the time of their applications.}
    \label{fig:majors}
  \end{center}
\end{figure}

Five hundred and thirty-four (534) candidates from fifty (50) countries applied for ASP2024 as shown in Figure~\ref{fig:applications}. At the time of writing, it is the largest number of applications received all the edition of ASP. 
All the applications went through a rigorous review process (by an international selection committee of twenty-five members). Applicants had to submit their curriculum vitae, university transcripts, a letter of motivation and arrange for one letter of recommendation. The selection committee (of 25 reviewers) was subdivided into eight subcommittees assigned to review a subset of the applications according to selection criteria set by the International Organizing Committee (IOC), considering requirements from funding agencies. Members of the selection committee were volunteers from the IOC, LOC adn lecturers. The distribution of selected students are shown in Figure~\ref{fig:selections}.
The female-to-male ratio of the selected students was greater than 7:10. In the in-person participants, there were more female students than male. Of the selected students, there were just a few declinations. Some students (and also lecturers) had issues in obtaining entrance visa for Morocco and could not attend in person. The selected students were required to have a minimum of three-year university education in engineering, computing, and fundamental and applied physics. Backgrounds on the selected students are shown in Figures~\ref{fig:degrees} and~\ref{fig:majors}.

\section{Expenditures}
\label{sec:exp}
The financial support received and described in Section~\ref{sec:sup} was used to cover expenses for ASP2024, with some reserve earmarked for ACP2025~\cite{ACP2025}. These include travels and full room and board for in-person students, local transportation, small detector lab equipment for students, teachers, and pupils.

\section{Feedback}
\label{sec:feed}
Towards the end of the event, we asked participants for feedback. This was carried out in two different approaches. In the first, in-person students were randomly arranged into eight groups to discuss within their groups and present collective feedback. The second approach was an anonymous survey where participants provided feedback on predefined survey questions.
\subsection{Collective feedback from student groups}
The different groups presented their feedback in a dedicated session on July 19, 2024. From these presentations, we noted the following positive points:
\begin{itemize}
\item ASP2024 offered high quality and passionate lecturers.
\item The organizing team demonstrated great flexibility and adaptability to the needs and suggestions of the participants.
\item There were good interactions between students, lecturers and organizers.
\item ASP2024 offered enhanced skills in the lab and networking opportunities with lecturers and other students.
\item Overall, it was a great and unforgettable experience, during which students learned a lot about physics and its applications, and expanded their networks of friendship and collaboration.
\end{itemize}
There were also many areas where the student groups suggested improvements. We note here the salient ones, to be addressed in future events:
\begin{itemize}
    \item Expand hands-on activities to more areas of physics and applications.
    \item Optimize the range and depth of lectures---plenary and parallel sessions, avoiding repetitive materials---while focusing on the research fields of the participants (and the host country). Preparing a summary of course contents well before the event may help in this optimization.
    \item Do more tests on the program logistics well before the event starts and protection against unsolicited online connections.
    \item Further optimize in the daily programs (coverage and duration), adding practical sessions and group projects to enhance understanding.
    \item Extend the program with field trips, visits to medical or research facilities in the area, and courses on scientific entrepreneurship. 
    \item Improve interactions of online participants.
\end{itemize}

\subsection{Feedback from anonymous survey}
\begin{figure}[!htbp]
 \begin{center}
  \includegraphics[width=\textwidth]{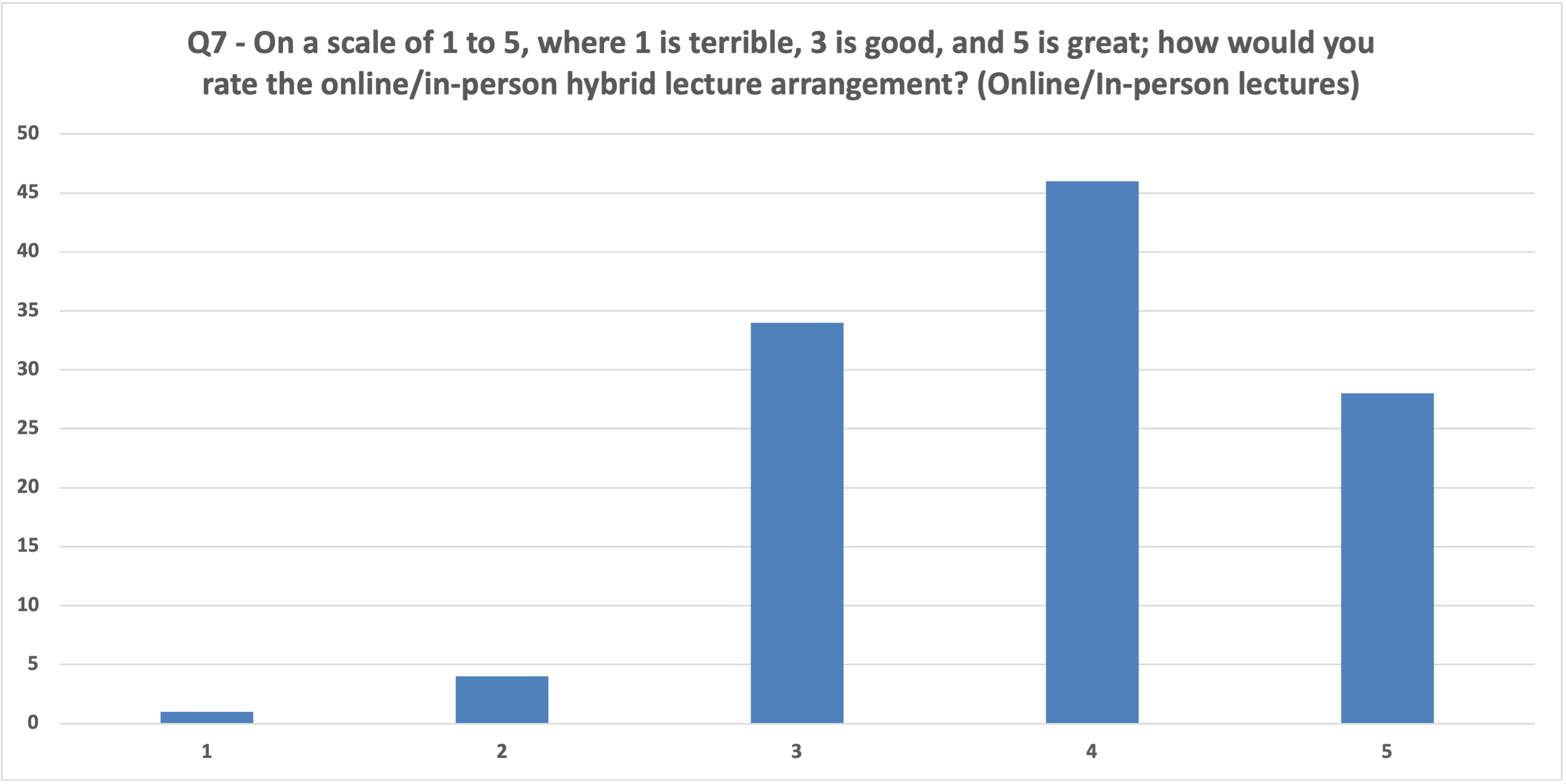}
   \caption{Participants' feedback on the hybrid arrangements.}
    \label{fig:hybrid}
  \end{center}
\end{figure}

\begin{figure}[!htbp]
 \begin{center}
  \includegraphics[width=\textwidth]{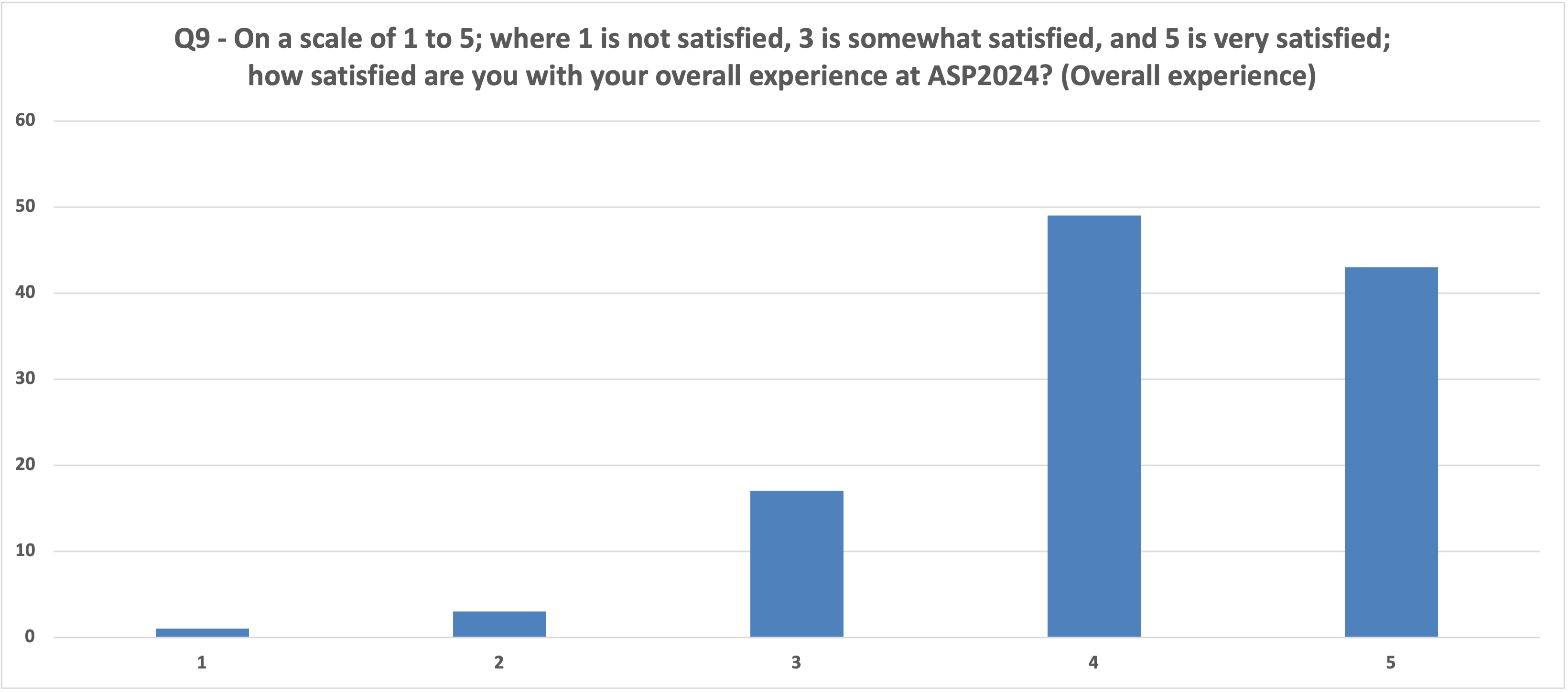}
   \caption{Participants were generally satisfied with the experiences at ASP2024.}
    \label{fig:satisfaction}
  \end{center}
\end{figure}

\begin{figure}[!htbp]
 \begin{center}
  \includegraphics[width=\textwidth]{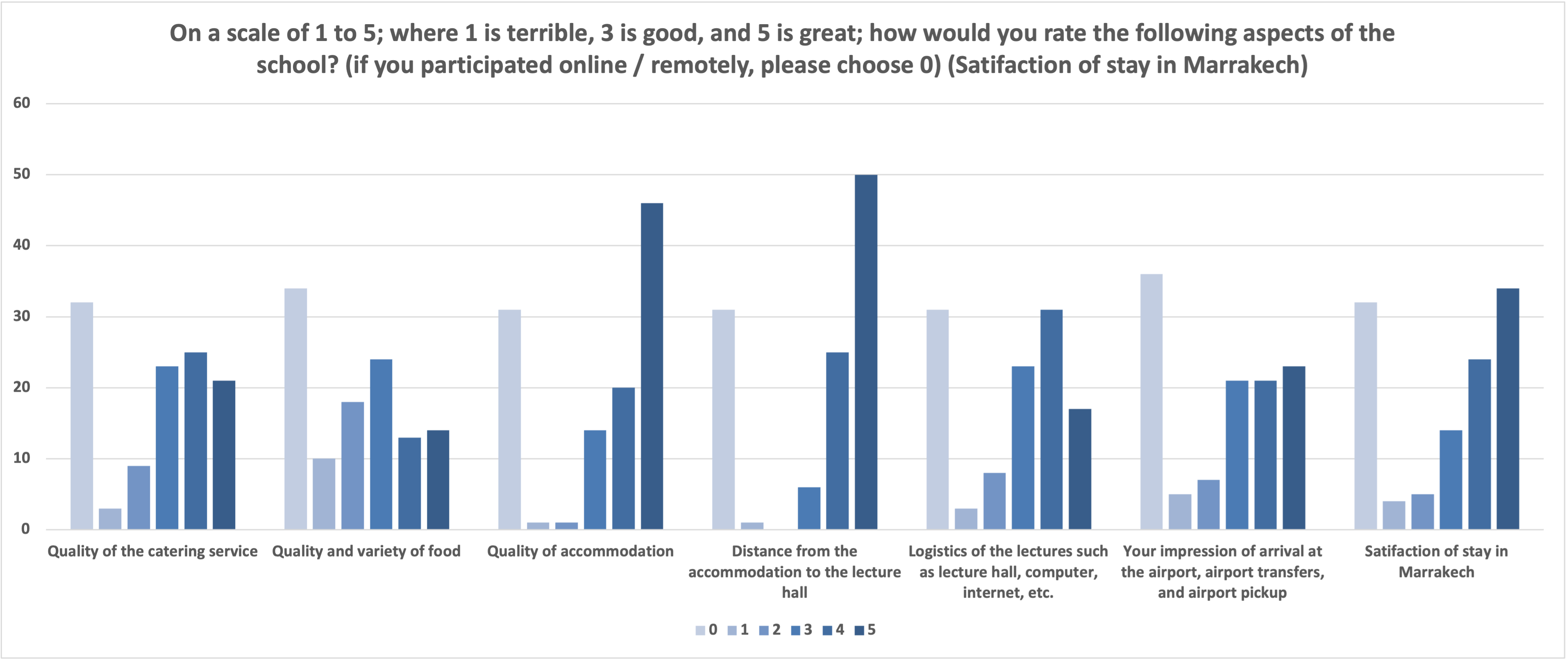}
   \caption{Feedback on various aspects of the logistics.}
    \label{fig:aspects}
  \end{center}
\end{figure}

\begin{figure}[!htbp]
 \begin{center}
  \includegraphics[width=\textwidth]{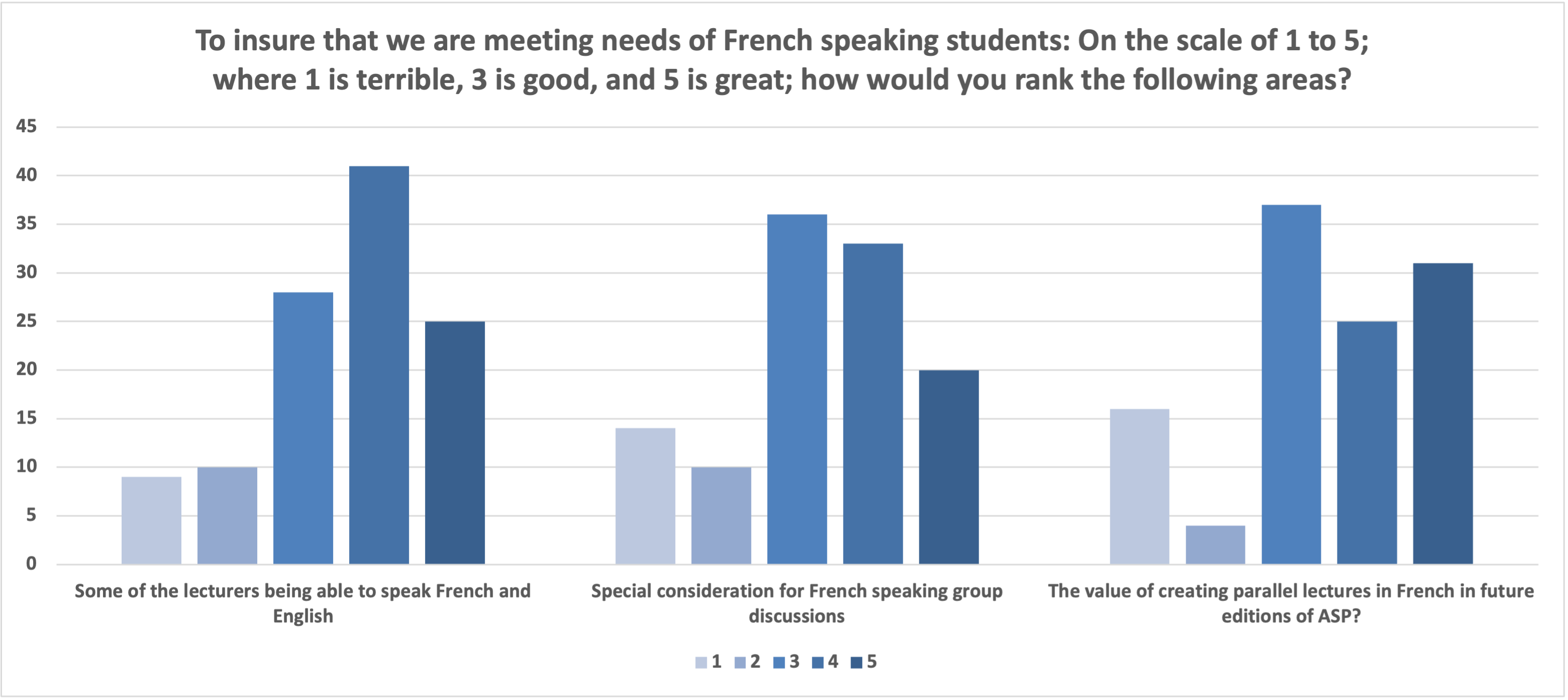}
   \caption{Feedback on the language of engagement. Consideration for French speaking participants was generally preferred.}
    \label{fig:language}
  \end{center}
\end{figure}

\begin{figure}[!htbp]
 \begin{center}
  \includegraphics[width=0.5\textwidth]{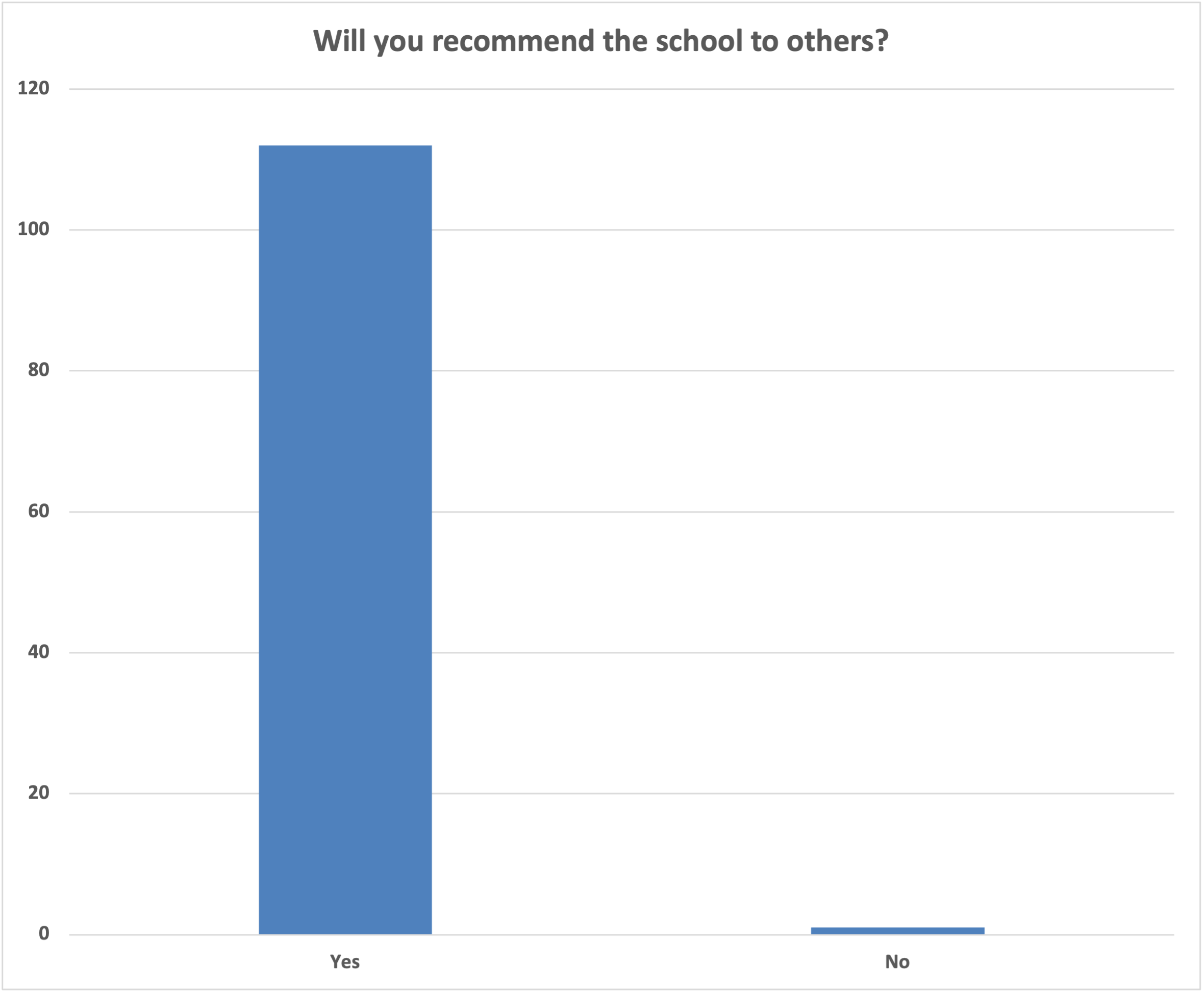}
  \includegraphics[width=0.45\textwidth]{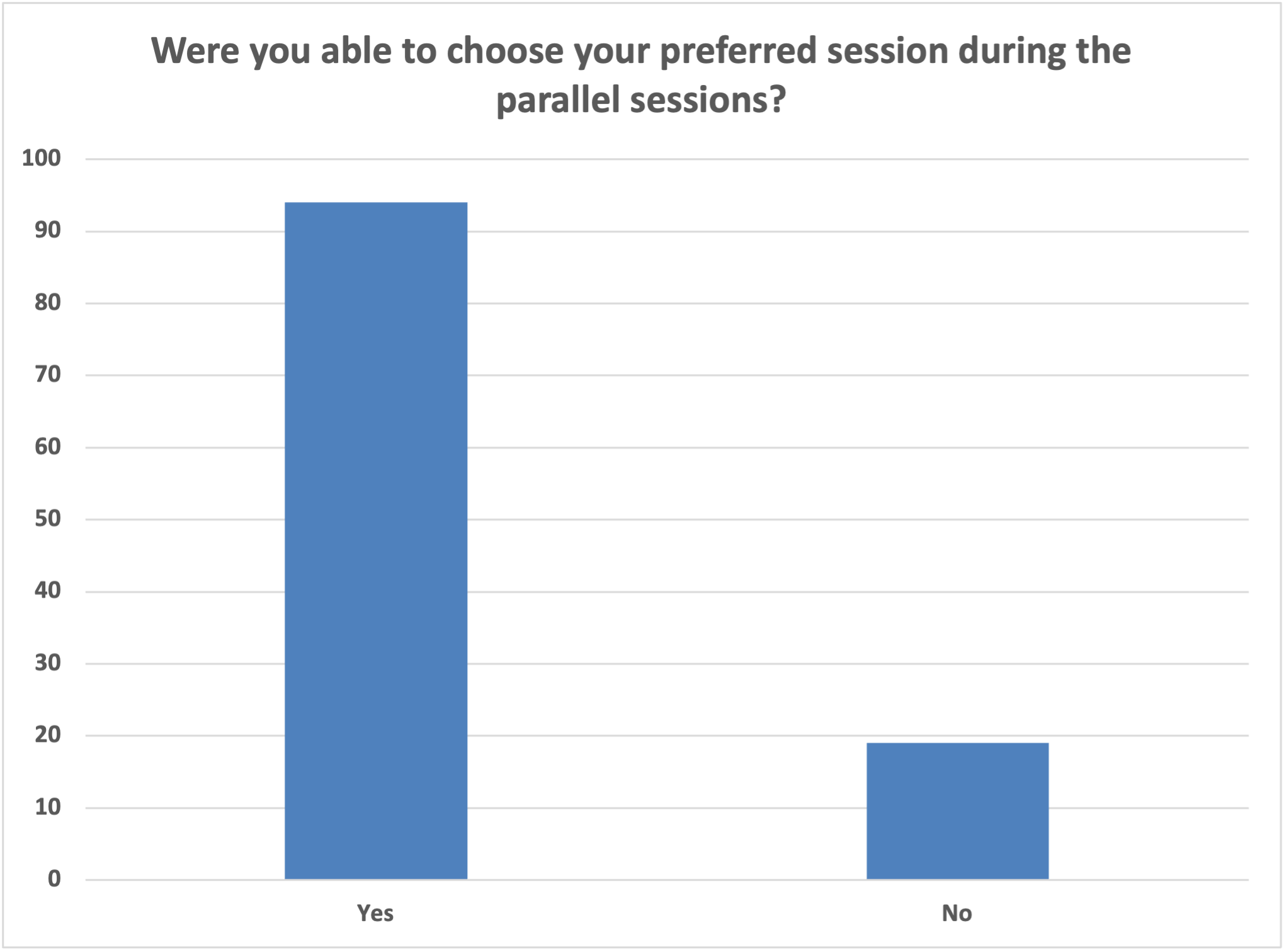}
  \includegraphics[width=0.45\textwidth]{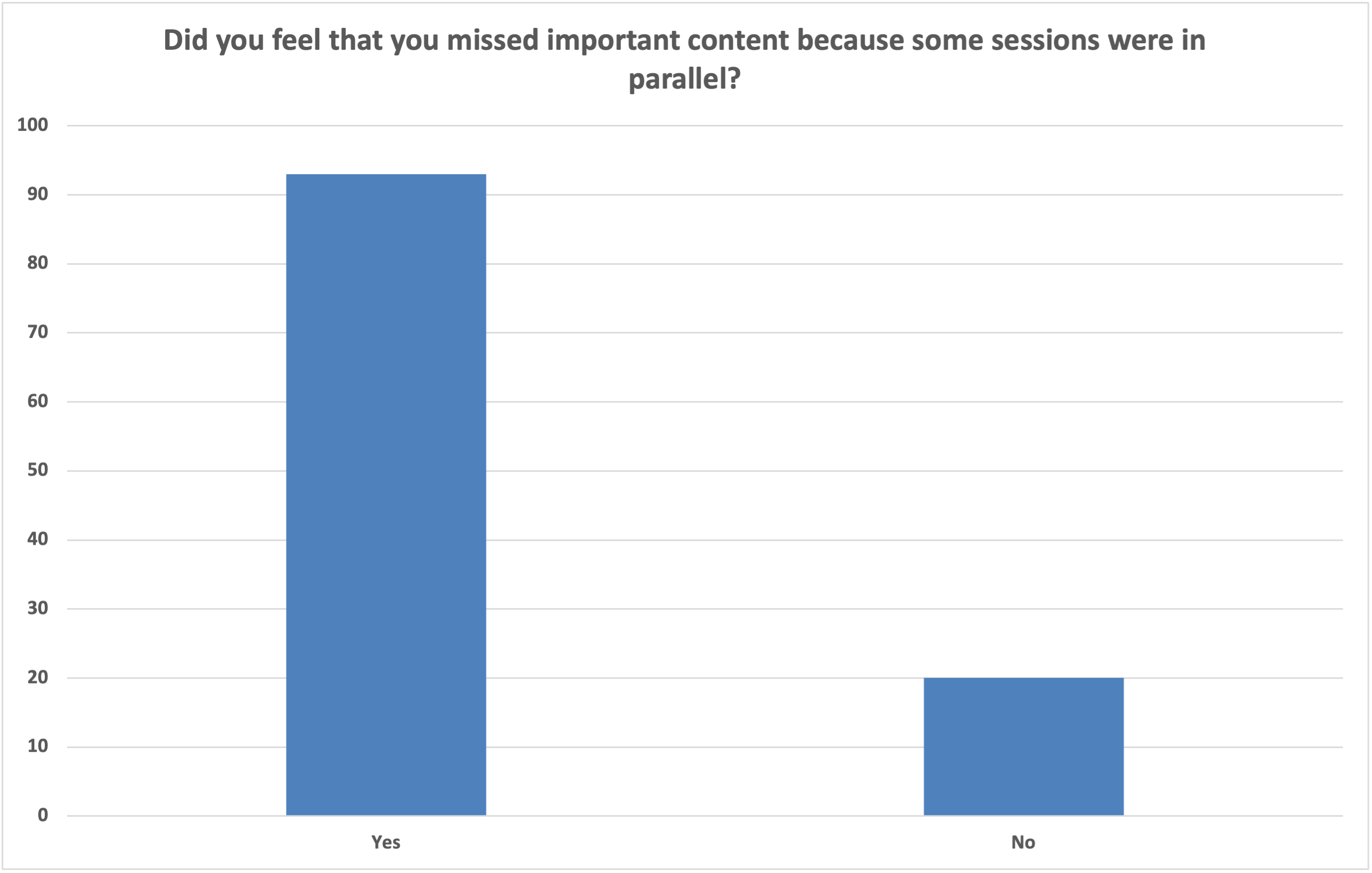}
  \includegraphics[width=0.45\textwidth]{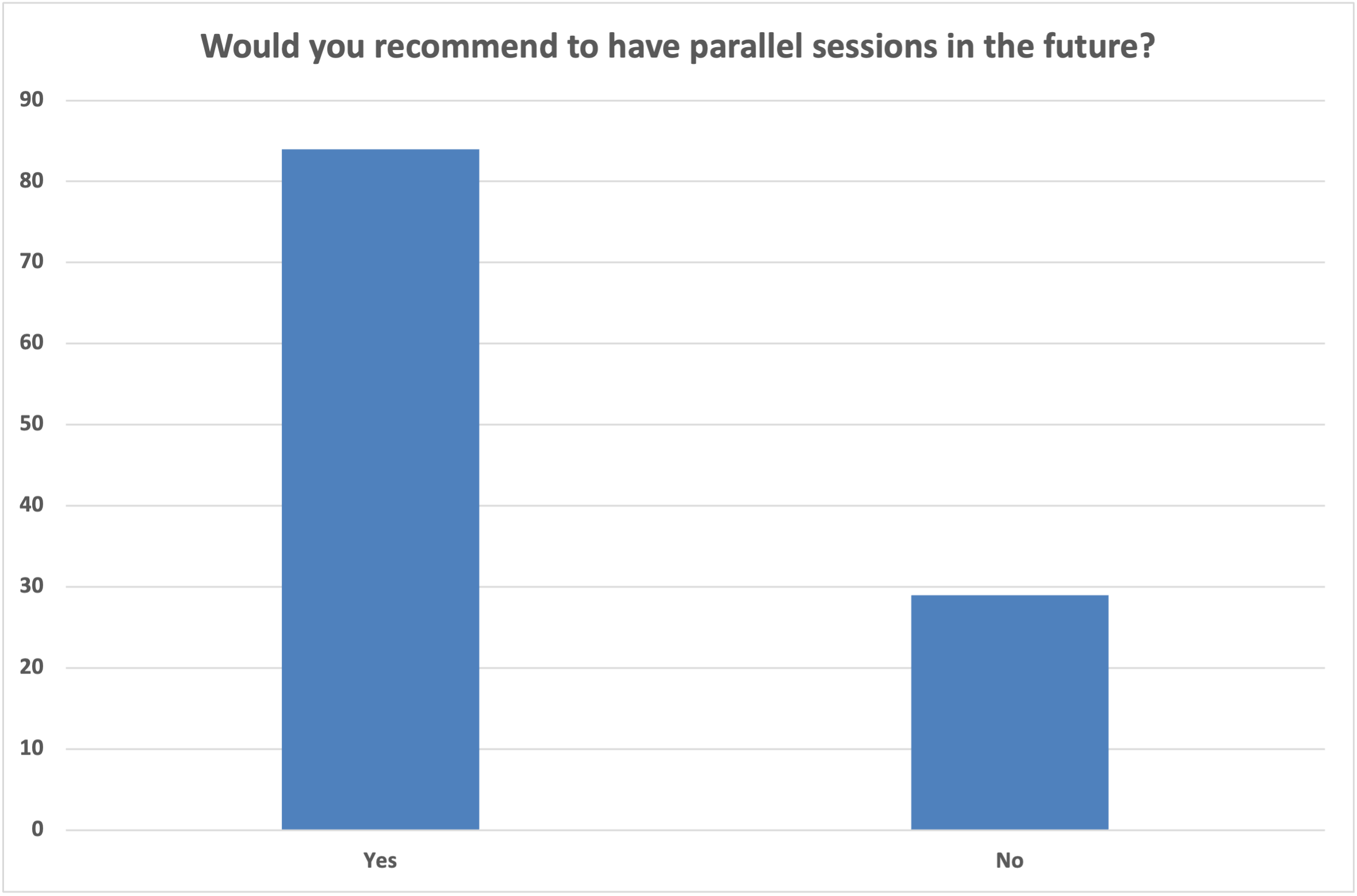}
   \caption{Most participants will recommend ASP to colleagues. However, parallel activities should be improved.}
    \label{fig:recommendation}
  \end{center}
\end{figure}

\begin{figure}[!htbp]
 \begin{center}
  \includegraphics[width=\textwidth]{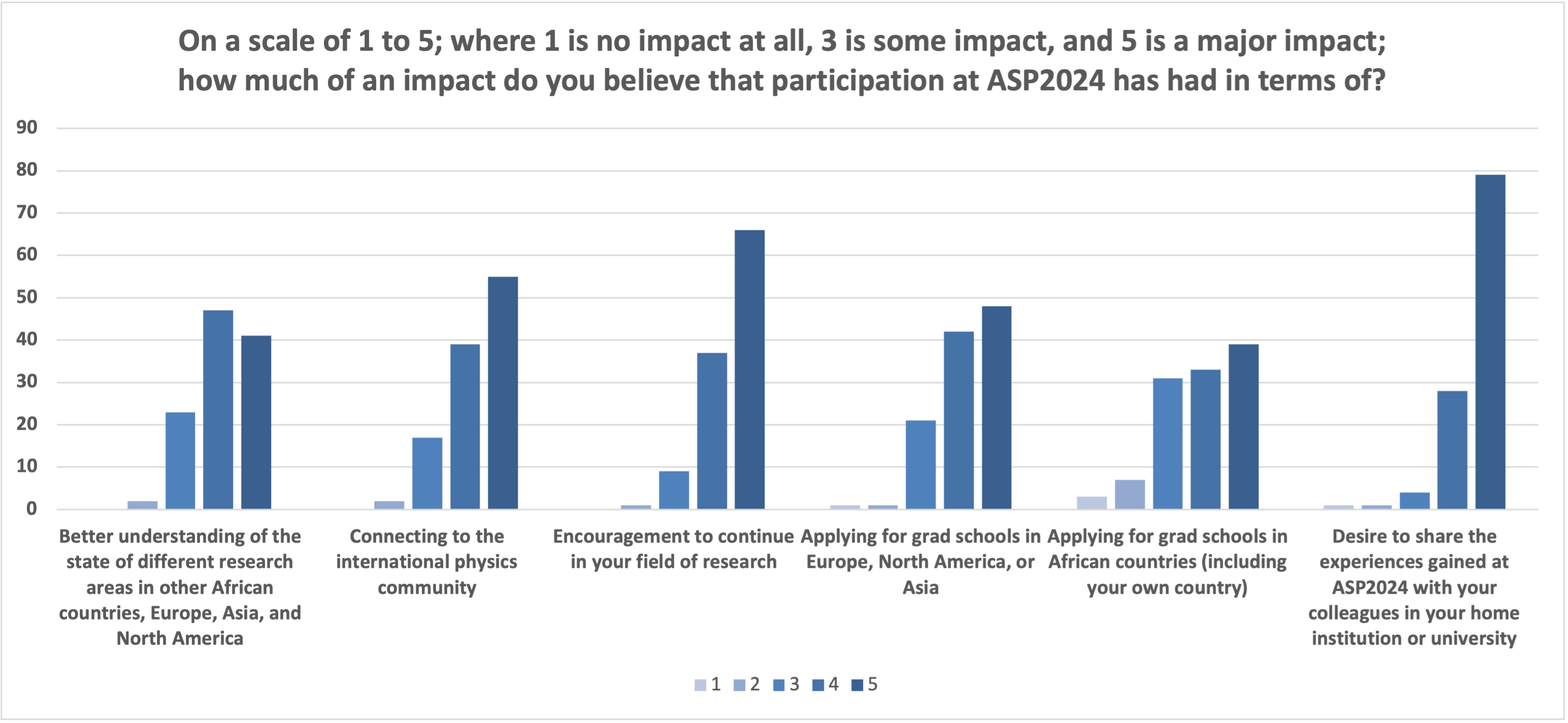}
   \caption{ASP2024 had positive impacts in various aspects of academic engagement.}
    \label{fig:impacts}
  \end{center}
\end{figure}

\begin{figure}[!htbp]
 \begin{center}
  \includegraphics[width=\textwidth]{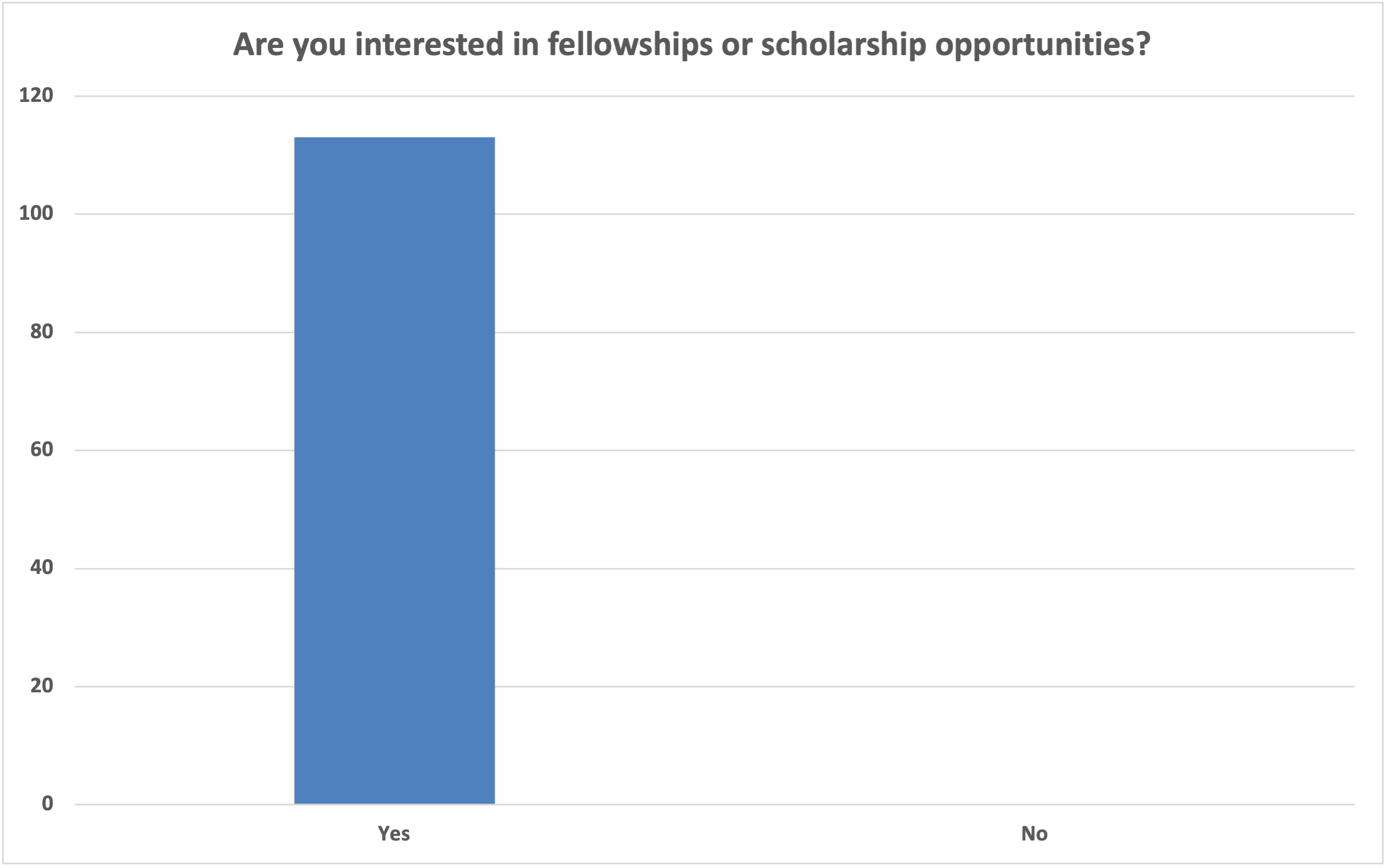}
   \caption{Participants expressed interest in higher education opportunities.}
    \label{fig:fellowships}
  \end{center}
\end{figure}

\begin{figure}[!htbp]
 \begin{center}
  \includegraphics[width=\textwidth]{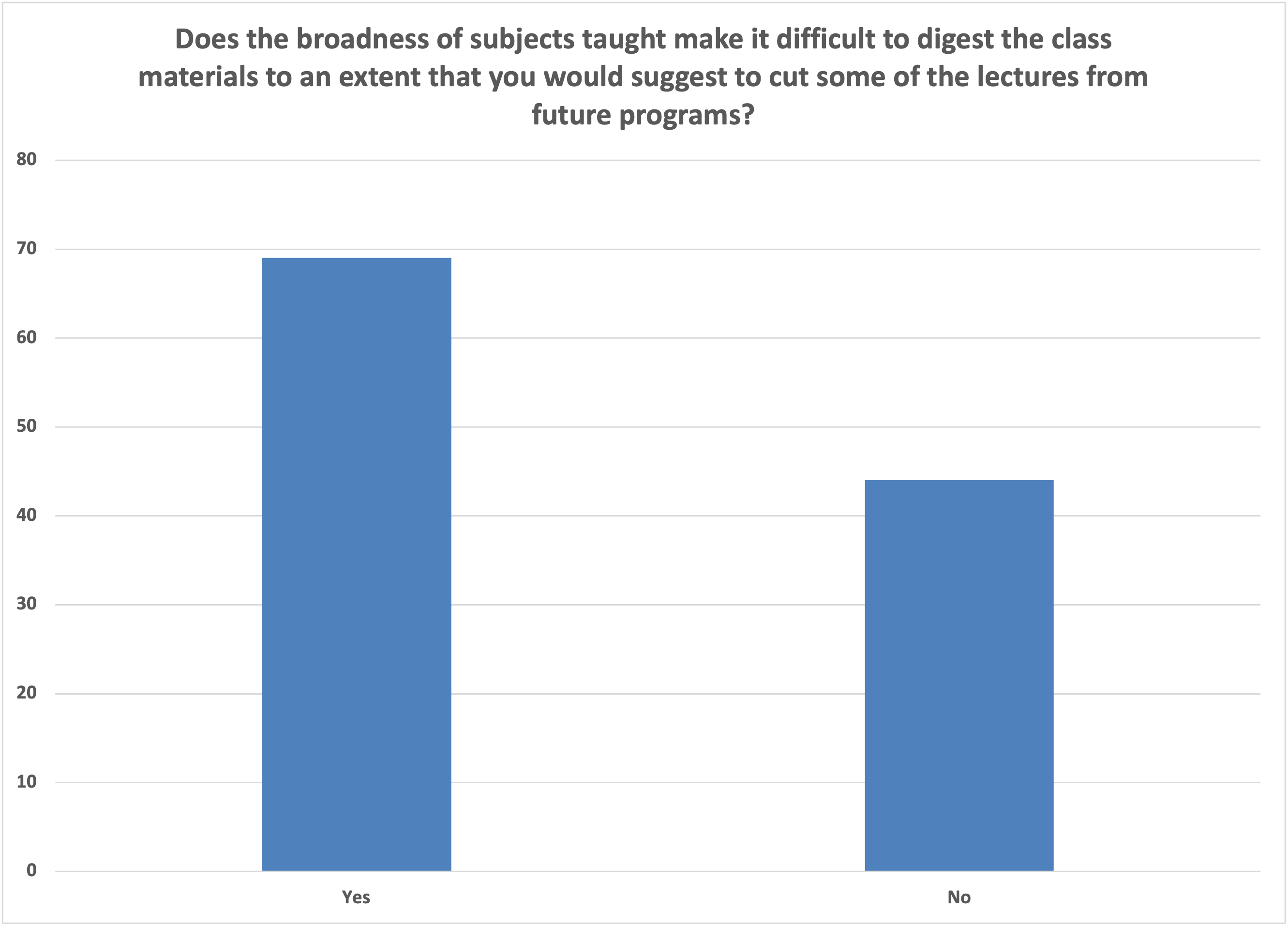}
   \caption{The broadness of the subjects taught posed difficulties for some participants.}
    \label{fig:digest}
  \end{center}
\end{figure}

\begin{figure}[!htbp]
 \begin{center}
  \includegraphics[width=\textwidth]{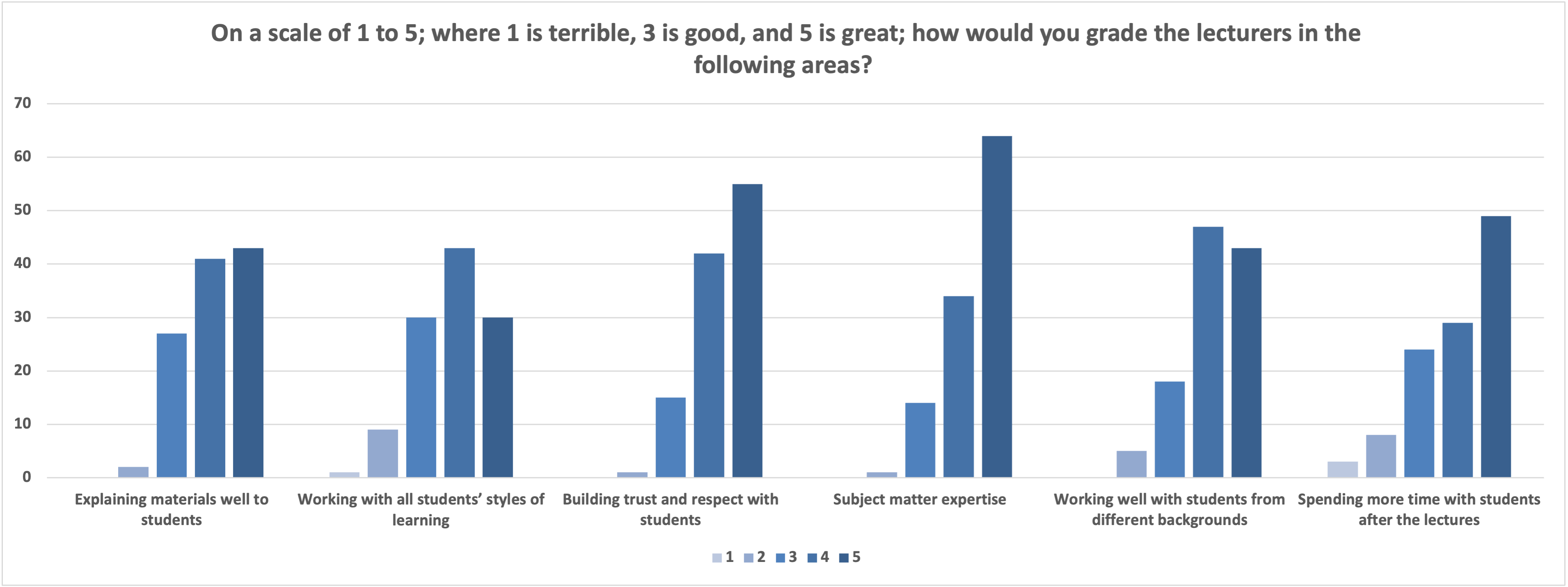}
   \caption{Participants reported positive interactions with lecturers.}
    \label{fig:lecturers}
  \end{center}
\end{figure}

\begin{figure}[!htbp]
 \begin{center}
  \includegraphics[width=\textwidth]{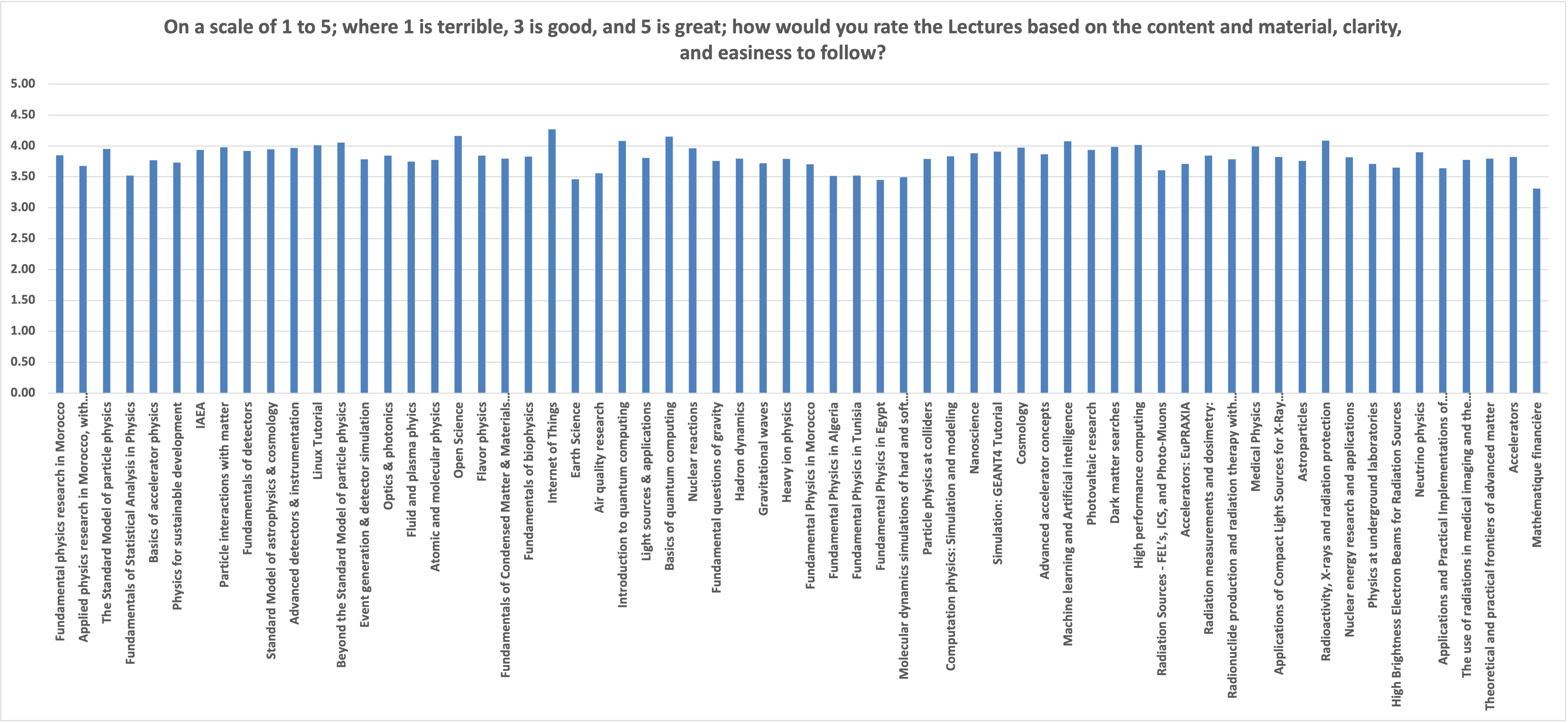}
   \caption{Participants rated favorably the lectures that they attended.}
    \label{fig:clarity}
  \end{center}
\end{figure}

\begin{figure}[!htbp]
 \begin{center}
  \includegraphics[width=\textwidth]{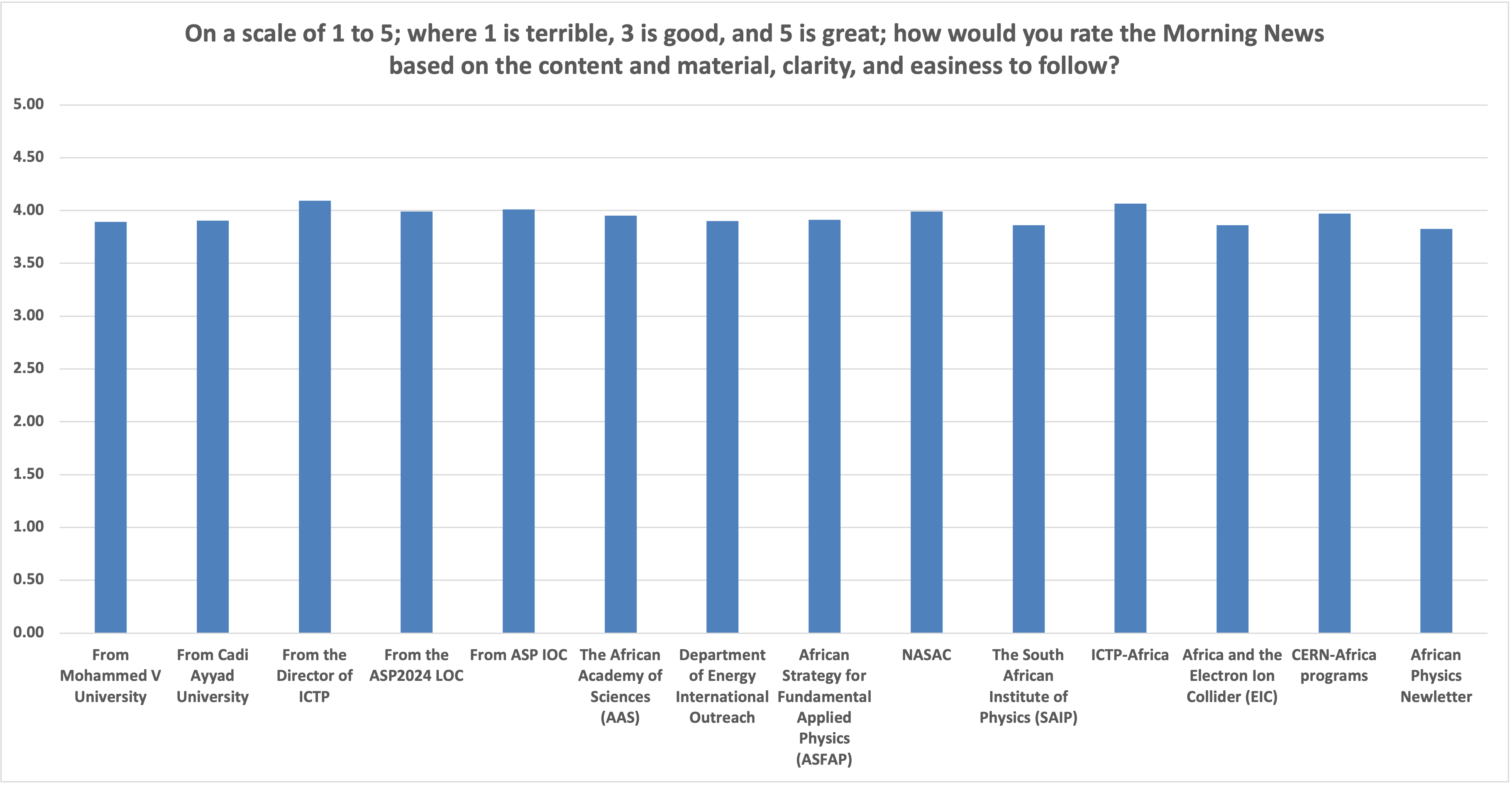}
   \caption{Participants' feedback on the early morning sessions.}
    \label{fig:morning}
  \end{center}
\end{figure}

\begin{figure}[!htbp]
 \begin{center}
  \includegraphics[width=\textwidth]{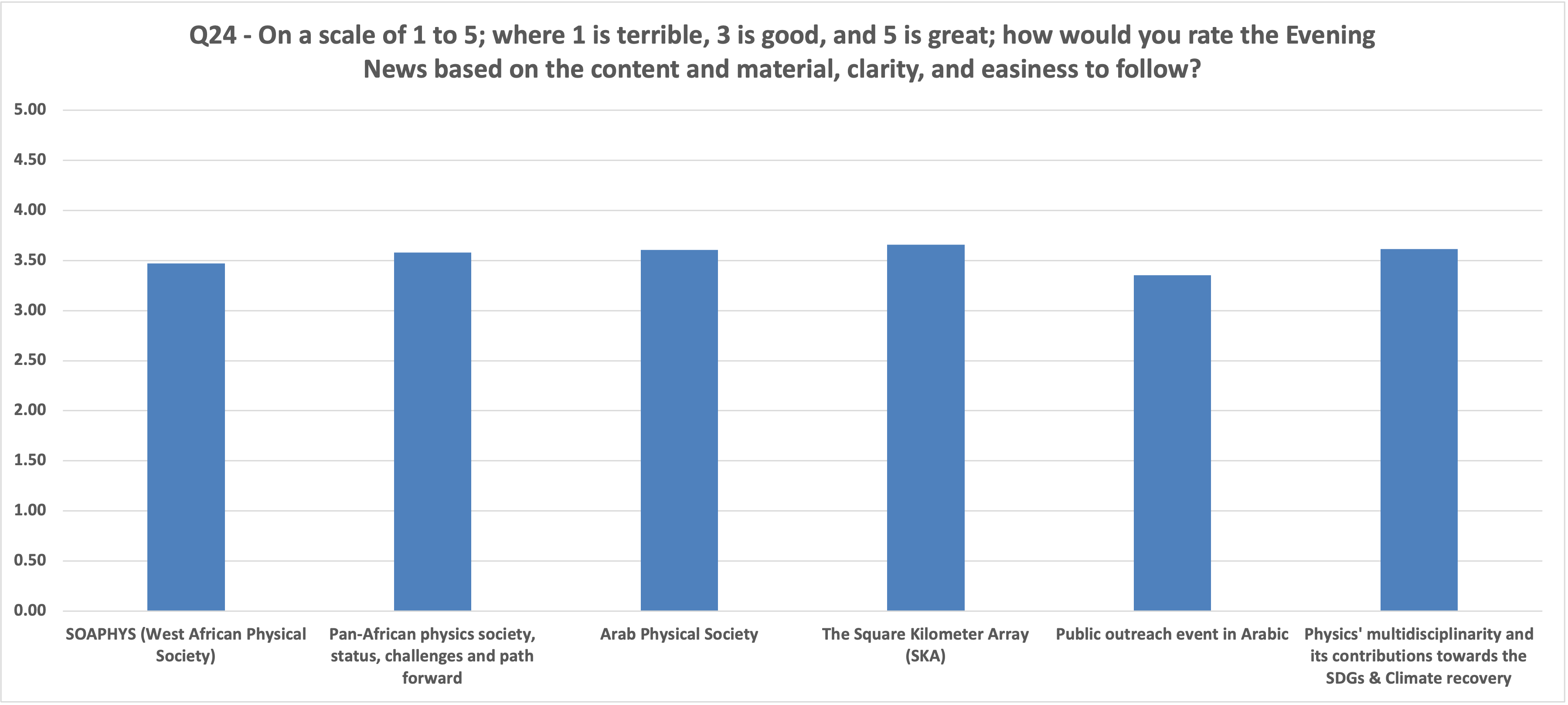}
   \caption{Participants' feedback on the evening sessions sessions.}
    \label{fig:evening}
  \end{center}
\end{figure}

\begin{figure}[!htbp]
 \begin{center}
  \includegraphics[width=\textwidth]{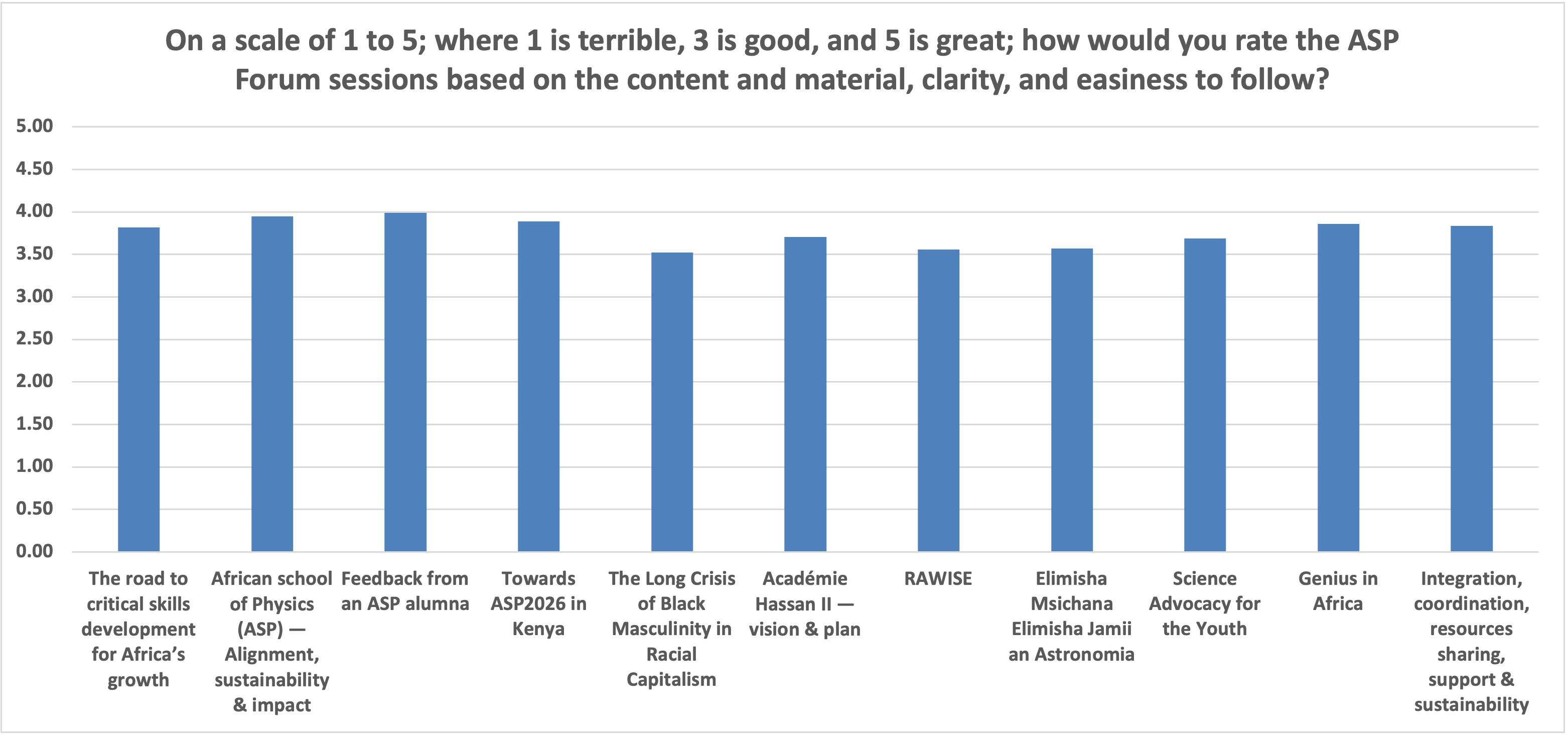}
   \caption{Participants' feedback on the ASP Forum.}
    \label{fig:forum}
  \end{center}
\end{figure}

One hundred-and-ten participants (in person and online) took the survey. Figure~\ref{fig:hybrid} shows that participants appreciated the hybrid arrangement, although online participants were less satisfied; understandably, most online participants expressed desire for in-person participation.
As shown in Figure~\ref{fig:satisfaction}, most of participants were satisfied with their ASP2024 experiences.
In the details, participants were satisfied with various aspects of logistics, as shown in Figure~\ref{fig:aspects}. However, the feedback shows that the quality of the catering services and diversity of meals could have been better.
In feedback presented in Figure~\ref{fig:language}, diversity in the languages of instructions and engagement should be entertained in future ASP considering the backgrounds of the participants.
Most survey respondents said they will recommend ASP to colleagues; however, improvements are needed in the organization of the parallel activities, as shown in Figure~\ref{fig:recommendation}.
In Figure~\ref{fig:impacts}, participants reported that ASP2024 had positive impacts on various aspects of their academic efforts.
Most participants were interested in fellowship opportunities for higher education, as shown in Figure~\ref{fig:fellowships}. 
Many participants reported that the broadness of the subjects presented made it difficult to digest lecture materials, as shown in Figure~\ref{fig:digest}; this issue is further discussed in Section~\ref{sec:out}.
In Figure~\ref{fig:lecturers}, participants were satisfied with their interactions with lecturers.
For the lectures that they did follow, participants rated favorably the content, material, clarity and easiness to follow, as shown in Figure~\ref{fig:clarity}.
Figures~\ref{fig:morning}, ~\ref{fig:evening} and ~\ref{fig:forum} show the feedback from students on the early morning and evening sessions about special topics, and on the ASP forum. General observations for the student survey results are as follows:
\begin{itemize}
\item As expected, the vast majority of survey respondents are from Morocco, going to Moroccan universities.
\item The vast majority of students are studying in their home countries.
\item About half of respondents will graduate in 2024 or next, and the majority are pursuing PhD.
\item About two-thirds respondents were in-person
students and were happy with the hybrid on-line format but there is room for improvement.
\item Similar results for overall satisfaction of ASP
with respect to logistics; the weakest part was the catering and the strongest was the accommodation.
\item There is room for improvement for accommodating French-speaking students.
\item The vast majority of students got the desired parallel sessions they wanted, but also felt they missed out on important materials. However, they still support parallel sessions moving forward.
\item Only 1 respondent would not recommend the school to others.
\item The biggest impact was in wanting to share experiences of ASP with colleagues at home institution and encouragement to continue in their area of research. However, students felt more strongly encouraged to apply to graduate schools in universities outside Africa instead of within, so there is some potential for improvement in growing connections/networking in Africa.
\item Everyone is interested in information on fellowships or scholarships, so the alumni network may consider continuing to expose students to these opportunities.
\item Lecturers were rated highly on subject matter expertise and building trust and report with students, but had the most room to grow on accommodating different learning styles.
\item The broadness/load of all the classes made the overall school difficult enough that over 60\% of the students would suggest cutting down the program in future editions.
\item The Morning News sessions were rated very highly overall, and had the best average score of any category (Morning News, Evening News, Lectures, Forum Sessions). The Evening News had the lowest average score.
\item The average scores for each lecture was generally high, but only nine sessions had an average score from participants of greater than four:
    \begin{itemize}
       \item Linux Tutorial.
       \item Beyond the Standard Model of particle physics.
       \item Open Science.
       \item Internet of Things.
       \item Introduction to quantum computing.
       \item Basics of quantum computing.
       \item Machine learning and artificial intelligence.
       \item High performance computing.
       \item Radioactivity, X-rays and radiation protection.
    \end{itemize}
\end{itemize}

%The articles of Ref.~\cite{ASP2022-articles} offer more feedback based on interviews and discussions during ASP2024.

\begin{figure}[!htbp]
 \begin{center}
  \includegraphics[width=\textwidth]{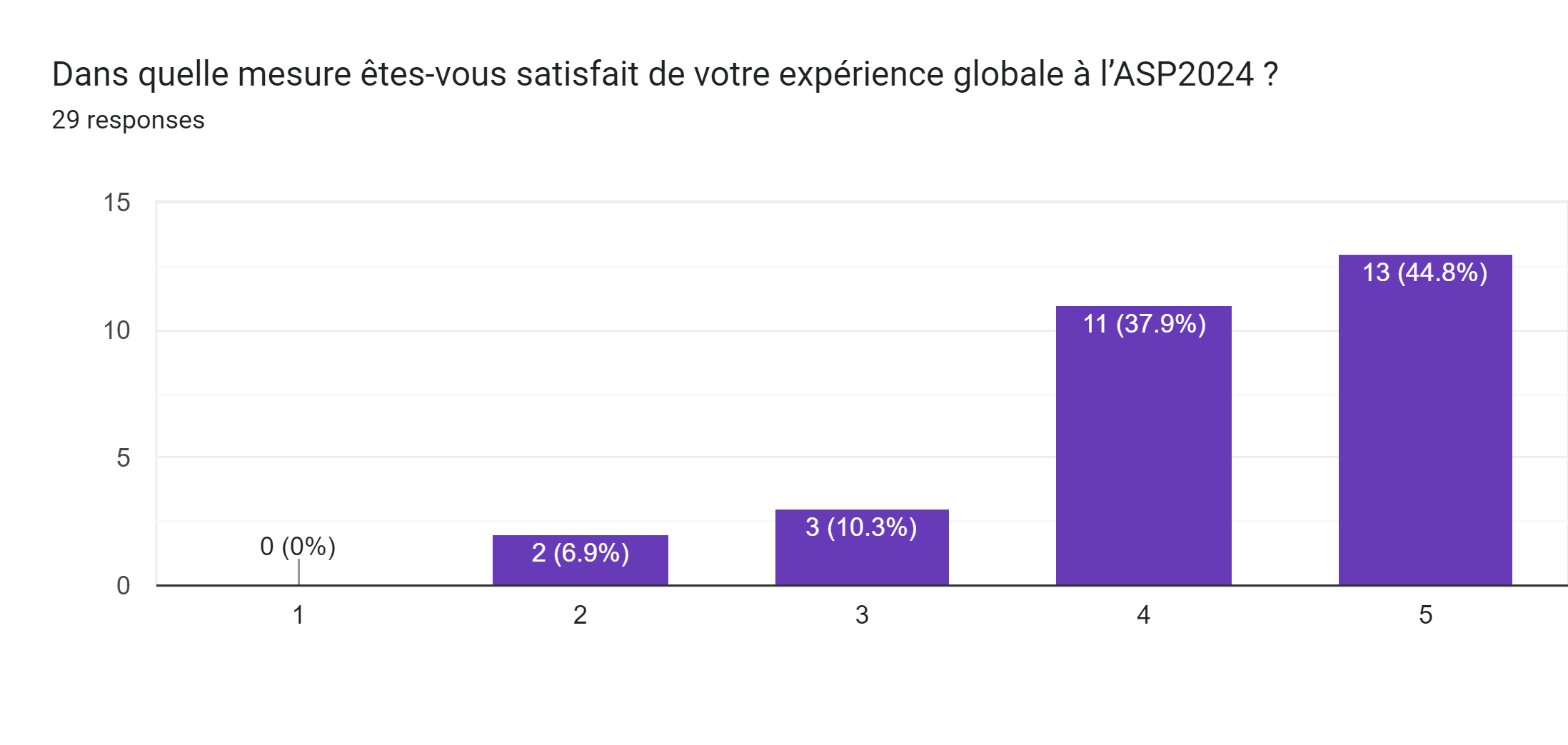}
   \caption{Feedback from high school teachers.}
    \label{fig:teachers-survey}
  \end{center}
\end{figure}

Another survey was conducted for the high school teachers who appreciated their plenary lectures on mechanic, optics \& photonics and thermodynamics. The teachers also commented favorably on the parallel hands-on activities in computing, Internet of Things and particle physics workshop---most found the sessions at the appropriate level and useful to their own teaching. Most the of teachers who responded to the survey said that they were satisfied with their experience at ASP2024, as shown in Figure~\ref{fig:teachers-survey}.

\section{Outlook}
\label{sec:out}

Going forward, improvements will be implemented to address feedback from the ASP2024 participants, as presented in Section~\ref{sec:feed}, to the extend feasible.

We look forward to organizing the fourth African Conference on Fundamental and Applied Physics (ACP2025) in Togo on September 14--20, 2025~\cite{ACP2025}. It will be an international physics conference, with strong participation of ASP alumni and African research faculties to present and discuss their research activities, and form new collaborations. ACP2025 is an activity of ASP, supported through the same funding cycles as the other activities mention in Section~\ref{sec:prog}.

The ninth African School of Physics, ASP2026, is planned in 2026 in Kenya.

\section{Conclusions}
\label{sec:conc}
ASP2024 was organized on April 15--19, 2024 (learners program) and July 7-21, 2024, at Cadi Ayyad University in Marrakesh, Morocco. It was as a hybrid event. In terms of the number of participants, ASP2024 was the largest ASP event to-date: there were over one thousand high school students, eighty high school teachers and 534 student applications, of which about 200 were selected; they were supported by many lecturers and organizers. ASP2024 learners' program was organized in April 2024 to maximize learners' attendance. The rest of the program was carried out in July 2024 to maximize the participation of students, teachers and lecturers. The scientific program was optimized in a combination of plenary and parallel activities to support the academic growth of the participants. The program was complemented with engagement with policymakers to discuss topics relevant to career development for African students. 

Feedback from participants was generally positive. We noted all the aspects to be improved in future events, namely ACP2025 and ASP2026 as mentioned in Section~\ref{sec:out}. 

\section*{Acknowledgments}
We thank all the sponsors of ASP2024 and the institutes who supported travels for lecturers. We also thank the organizing committee (local and international), in particular Brian Masara (SAIP), Gilbert T?(C)kout?(C) (Executive Management School of Paris, France), Abdelkarim Boskri and Abdelmonaime Elmaaoua (Cadi Ayyad University), Sanae Samsam (INFN-Milan), and Rajaa Sebihi (Mohammed V University). We appreciate the efforts of Viktoriya Lvova (ICTP, SMR3955 -- Secretariat) with the management of student application data and arrangement of international travels. We are grateful for the tremendous efforts of the lecturers. We acknowledge the collegial atmosphere maintained by all the participants, i.e. students, teachers, lecturers, pupils, organizers and policymakers.

\newpage

\bibliographystyle{elsarticle-num}
\bibliography{myreferences} 

\begin{thebibliography}{10}
\expandafter\ifx\csname url\endcsname\relax
  \def\url#1{\texttt{#1}}\fi
\expandafter\ifx\csname urlprefix\endcsname\relax\def\urlprefix{URL }\fi
\expandafter\ifx\csname href\endcsname\relax
  \def\href#1#2{#2} \def\path#1{#1}\fi

\bibitem{ASP2021-reports}
{Kétévi A. Assamagan, Bobby Acharya, Temitope Adenuga, Mohamed Chabab, Kenneth Cecire, Simon H. Connell, Anne E. Dabrowski, Christine Darve, Farida Fassi, Jonathan R. Ellis, Fernando Ferroni, Mounia Laassiri, Steve G. Muanza}, {Activity report of the African School of Physics, 2019-2021}, {arXiv:2109.00509} ({2019--2021}).
\newblock \href {https://doi.org/https://doi.org/10.48550/arXiv.2109.00509} {\path{doi:https://doi.org/10.48550/arXiv.2109.00509}}.

\bibitem{ASP}
{B. S. Acharya, K. A. Assamagan, A. E. Dabrowski, C. Darve, J. Ellis, S. Muanza}, {The African School of Physics}, \url{https://www.africanschoolofphysics.org/}.

\bibitem{ASP-reports}
{Activity reports of African School of Physics}, \url{http://africanschoolofphysics.web.cern.ch/2010/asp2010.pdf, https://africanschoolofphysics.web.cern.ch/asp2012/asp2012_final.pdf, https://www.africanschoolofphysics.org/wp-content/uploads/2014/11/asp2014.pdf, https://www.africanschoolofphysics.org/wp-content/uploads/2019/08/ASP2016-FinalReport.pdf, https://www.africanschoolofphysics.org/wp-content/uploads/2019/08/ASP2018.pdf} (2010-2018).

\bibitem{asp2018}
{B. S. Acharya, K. A. Assamagan, M. Backes, K. Cecire, A. E. Dabrowski, C. Darve, J. Ellis, J. A. Gray, E. Kasai, S. Muanza, J. Ndjamba1, A. Philander, M. Shahungu, G. Simon, D. Singh, R. Steenkamp, R. Voss, A. Zulu}, {Activity Report on the Fifth Biennial African School of Fundamental Physics and Applications}, \url{https://www.africanschoolofphysics.org/wp-content/uploads/2019/08/ASP2018.pdf} (2018).

\bibitem{ASP2022}
K.~A. Assamagan, B.~Acharya, K.~Cecire, C.~Darve, F.~Ferroni, J.~A. Gray, A.~Muronga, \href{https://arxiv.org/abs/2302.13940}{Activity report on the seventh african school of fundamental physics and applications (asp2022)} (2023).
\newblock \href {http://arxiv.org/abs/2302.13940} {\path{arXiv:2302.13940}}.
\newline\urlprefix\url{https://arxiv.org/abs/2302.13940}

\bibitem{ACP2025}
{The 4th African Conference on Fundamental and Applied Physics, ACP2025}, \url{https://indico.cern.ch/event/1458227/} (September 14--20, 2025).

\bibitem{AfPS}
{The African Physical Society}, \url{https://www.africanphysicalsociety.org/}.

\bibitem{acp2021}
{K\'et\'evi A. Assamagan, Mohamed Chabab, Farida Fassi, Ulrich Goerlach, Mohamed Gouighri, et al.}, {The Second African Conference on Fundamental and Applied Physics}, \url{https://indico.cern.ch/event/1060503/} (2022).

\bibitem{ReportLearners2024}
K.~A. Assamagan, A.~Boskri, K.~Cecire, M.~Chabab, C.~Darve, F.~Fassi, M.~Laassiri, S.~Samsam, J.~Vischer, \href{https://arxiv.org/abs/2408.01464}{Summary report on the 2024 african school of physics program for learners} (2024).
\newblock \href {http://arxiv.org/abs/2408.01464} {\path{arXiv:2408.01464}}.
\newline\urlprefix\url{https://arxiv.org/abs/2408.01464}

\bibitem{Students2024}
{The ASP2024 scientific program for students}, \url{https://indico.cern.ch/event/1417241/timetable/?view=standard}.

\bibitem{Teachers2024}
{The ASP2024 scientific program for high school teachers}, \url{https://indico.cern.ch/event/1416196/timetable/?view=standard}.

\bibitem{Forum2024}
{The ASP2024 forum day program}, \url{https://indico.cern.ch/event/1417241/timetable/?view=standard\#day-2024-07-13}.

\bibitem{ICTP-ASP2024}
{Nurturing Young Physicists in Africa}, \url{https://www.ictp.it/news/2024/9/nurturing-young-physicists-africa}.

\bibitem{Learners2024}
{The ASP2024 scientific program for learners}, \url{https://indico.cern.ch/event/1393743/timetable/?view=standard}.

\end{thebibliography}

\end{document}